\documentclass[12pt]{iopart}
\usepackage{graphicx}
\usepackage{natbib}
\newcommand{\rmno}{RMn$_2$O$_5$ }
\newcommand{\R}{RMn$_2$O$_5$ }
\newcommand{\Y}{YMn$_2$O$_5$ }
\newcommand{\Bi}{BiMn$_2$O$_5$ }
\newcommand{\Ho}{HoMn$_2$O$_5$ }
\newcommand{\Dy}{DyMn$_2$O$_5$ }
\newcommand{\Tb}{TbMn$_2$O$_5$ }

\newcommand{\Tbun}{TbMnO$_3$ }
\newcommand {\Mt} {Mn$^{3+}$ }
\newcommand {\Mf} {Mn$^{4+}$ }
\newcommand {\DM} {Dzyaloshinskii-Moriya }

\begin{document}

\title[]{A Neutron diffraction study of multiferroics RMn$_2$O$_5$}

\author{P. G. Radaelli}
\address{ISIS Facility, Rutherford Appleton Laboratory - STFC, OX11 0QX, United Kingdom}
\author{L. C. Chapon}
\address{ISIS Facility, Rutherford Appleton Laboratory - STFC, OX11 0QX, United Kingdom}

\begin{abstract}
The magnetic properties of \R multiferrroics as obtained by unpolarized and polarized neutron diffraction experiments are reviewed. We discuss the qualitative features of the magnetic phase diagram both in zero magnetic field and in field and analyze the commensurate magnetic structure and its coupling to an applied electric field. The origin of ferrolectricity is discussed based on calculations of the ferroelectric polarization predicted by different microscopic coupling mechanisms (exchange striction and cycloidal spin-orbit models). A minimal model containing a small set of parameters is also presented in order to understand the propagation of the magnetic structure along the \textit{c}-direction.

\end{abstract}

\maketitle

\section{Introduction}

\indent The recent discovery of a new class of magnetic ferroelectric materials, in which electrical polarization coincides with a magnetic ordering or reordering transition, has generated very significant interest \cite{ISI:000248912900036,ISI:000239792700035,ISI:000221644600033,ISI:000186370800038,ISI:000254144700013, ISI:000243225400013}. The attractive feature of these materials is not so much the value of the electrical polarization, which is several orders of magnitude smaller than for typical ferroelectrics and even for "classic" magnetic multiferroics such as BiFeO$_3$, but rather the very large cross-coupling between magnetic and electrical properties, which makes the "novel" multiferroics into enticing paradigms of functional behavior.  Although conceptual devices based on novel multiferroics have been discussed \cite{ISI:000239792700035}, none of the materials so far described are directly suitable for applications, because the transition temperatures are low.  What keeps much of the interest alive, however, is the possibility of discovering an underpinning general principle which could be applied to guide the synthesis of bulk of film materials with better properties. Some basic common facts about these materials have been established with clarity.  As for conventional ferroelectrics, electrical polarization emerges as a result of a symmetry-breaking transition from a high-temperature paraelectric phase.  Unlike conventional ferroelectrics however, the primary order parameter for this phase transition is \emph{magnetic} rather than structural:  as a result, the overall magneto-structural symmetry is lowered from that of the paramagnetic phase, eventually leading to a polar point group below one of the magnetic transition temperatures.  Here, ferroelectricity is induced by some form of magneto-elastic and/or magneto-electronic interaction. On this principle, much work has been published on both group theory \cite{ISI:000249155100099} and phenomenology \cite{ISI:000249155100118,ISI:000237404200247,ISI:000235394100083}, aimed at establishing the symmetry requirements for the appearance of ferroelectricity, as well as the coupling between different order parameters and the generalized phase diagrams of these materials. Essentially all of these results are independent of the microscopic magneto-electric coupling mechanism. There is in fact no requirement that this mechanism is one and the same for all "novel" multiferroics  --- in this case, symmetry would be the single unifying principle.  However, establishing this microscopic mechanism in each case is crucial for a quantitative understanding on the phenomenon, and it is an essential guiding principle for engineering new materials.  As it happens, most of the "novel multiferroic" materials so far discovered share much more than a broken magnetic symmetry leading to a polar group:  in fact, the presence of a cycloidal component to the magnetic structure in \cite{ISI:000231310900063,ISI:000230276700081,ISI:000186370800038,ISI:000250620300074,ISI:000225661100073,ISI:000179285000001} suggests a common underlying microscopic mechanism, which is critically dependent on the non-collinearity of the spins.  Nagaosa \textit{et al.} \cite{ISI:000254311900014} and M. Mostovoy \cite{ISI:000235394100083} have explored this concept early on from different angles, and proposed what has now become known as the "theory of ferroelectricity in cycloidal magnets".  In particular, Nagaosa \textit{et al.} have proposed a detailed microscopic model based on the relativistic spin-orbit interaction, which is able to predict qualitatively, and to a certain extent quantitatively, the emergence of ferroelectricity in the presence of a cycloidal (or more generally non-collinear) magnetic structure of appropriate symmetry.  It has been known for a rather long time that other mechanisms can potentially induce ferroelectricity even in collinear antiferromagnets.  However, the appeal of a single "universal" model is strong, and physics based on cycloidal modulations is very often sought as "the" single explanation for novel multiferroic behavior.  The family of compounds with general formula \rmno ($R$=Y, Rare earth, Bi and La) has so far stubbornly resisted being classified with the other compounds in the "cycloidal multiferroic" family.  Three aspects of the \rmno phenomenology stand out to suggest that these materials may be in a class of their own:  the fact that the direction of the electrical polarization cannot be turned by an applied magnetic field \cite{ISI:000231564400088}, the \emph{commensurate} nature of the magnetic ferroelectric phase (ferroelectricity is all but lost in the low-temperature, incommensurate phase),  and the fact that in the ferroelectric phase, moments in the \textit{ab}-plane are almost collinear.  The idea that \rmno may stand apart form the other multiferroics has been recently challenged by new findings --- particularly the evidence for a previously unobserved cycloidal components \cite{ISI:000248244500035,vecchini}. In this paper, we describe the general phenomenology and present new results on the \rmno family  multiferroics.  We argue that, in spite of the emerging complexity and subtlety of the magnetic structure in these materials, there are still strong reasons to believe that cycloidal physics plays a minor role in \rmno (at least for the commensurate phase), and that ferroelectricity emerges due to more conventional exchange-striction effects in the context of a structure with built-in charge ordering.

\section{Experimental}

\indent Polycrystalline, single-phase \R samples were prepared through conventional solid-state reaction in an oxygen environment. Stoichiometric quantities of Tb$_4$O$_7$ purity
99.998\%\ , Dy$_2$O$_3$ 99.99\%\ , Ho$_2$O$_3$ 99.995\%\ , and MnO$_2$
 99.999\%\ were thoroughly mixed, compressed into pellets,
and then sintered at 1120 $^{\circ}$ C for 40 h with intermediate
grindings. The samples were finally cooled at 100 $^{\circ}$ C per
hour down to room temperature. Single crystals were prepared using the method described in \cite{vecchini}.
Powder neutron diffraction data were collected on the GEM diffractometer at the
ISIS facility of the Rutherford Appleton Laboratory (UK). The
samples were enclosed in vanadium cans. All the data presented have been collected
using either a helium cryostat or an Oxford Instrument 10 Tesla cryomagnet.
For zero-field experiment, data have always been collected on warming, after cooling the sample
to base temperature (typically 1.6K). Experiments performed in magnetic fields
were conducted as follows:
Meaurements on the \Y\ compound
were carried out at 1.6K in field of respectively 0,2,4,6,8 Tesla.
For each magnetic field, data were acquired for 2 hours.
Measurements on the \Tb\ system were carried out in the temperature
range 20 to 34 K, in steps of 1K, for several values of the magnetic
field (0,1,3,5,7,9 Tesla). For each measurement, the sample was
first zero-field cooled. The magnetic field was then applied and
measurement were carried out on warming, setting a counting time of
1 hour per temperature.
Single crystal neutron diffraction experiments have been recorded at the Institut Laue Langevin (France),
using the D10 4-circle diffractometer for unpolarized work
and the D3 instrument equipped with CRYOPAD for Spherical Neutron Polarimetry.
Details of the experimental procedures can be found in \cite{vecchini} and \cite{Radaelli_D3}.
Analysis of powder and single-crystal neutron diffraction data were carried out with the software FullProf\cite{FullProf}.

\section{Crystal structure}
\begin{figure}[]
\includegraphics[scale=1,angle=-90]{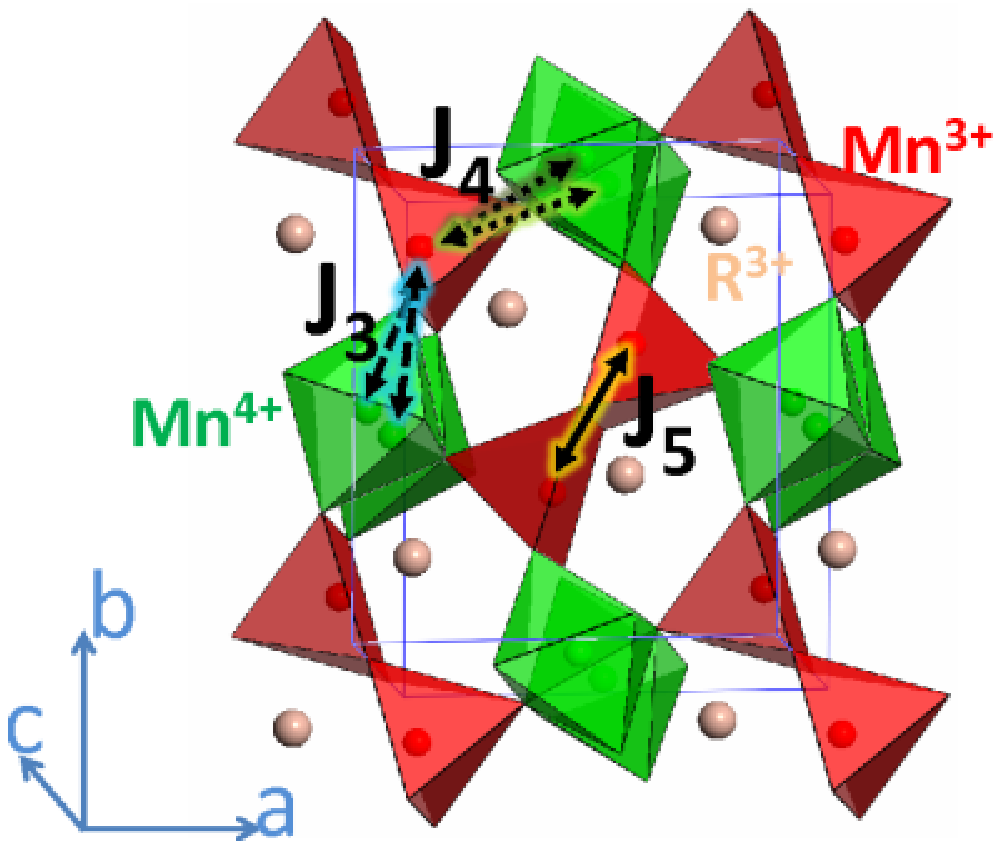}
\caption{Perseptive view of the crystal structure of \R. \Mt , \Mf and $R^{3+}$ ions are shown respectively as red, green and pink spheres. Polyhedra around \Mt and \Mf ions connecting first neighbour oxygens ions are also shown. Magnetic super-exchange interactions J$_i$, i=3,5 are shown by double side arrows.}
\label{nuclearstructure}
\end{figure}

\indent The crystal structure of \R compounds, already described in details in \cite{ISI:000230276600039,vecchini}, will be briefly reviewed here.
The Mn ions are fully charge ordered, with \Mt and \Mf ions occupying sites of different symmetry. \Mt and \Mf are respectively coordinated by five oxygens in square pyramid geometry and six oxygens in octahedral geometry, as shown in Figure \ref{nuclearstructure}. The crystal structure is best described by considering the \textit{ab}-plane configuration and out-of-plane configuration independently. In plane, octahedra and pyramids are corner-sharing through either the pyramid base or pyramid apex. In addition, adjacent pyramids are connected through their base. In total, three inequivalent exchange paths between magnetic ions exist in the plane (noted J$_3$, J$_4$ and J$_5$ in Fig. \ref{nuclearstructure} following notations of \cite{ISI:000230276600039}). Along the \textit{c}-axis, octahedral sites are sharing edges, forming linear chains. The \Mf are located at z$\sim$0.25 and (1-z)$\sim$0.75, so that \Mt ions, positioned at z=$\frac{1}{2}$, form layers in between adjacent \Mf. R$^{3+}$ ions occupy sites of same symmetry than \Mt in the z=0 plane, forming another layer alternating with the \Mt layer. The ions in the primitive unit cell are labelled following the convention defined in \cite{ISI:000230276600039,vecchini}, as indicated in Fig. \ref{nuclearstructure}.

\section{Magnetic ordering -sequence of phase transitions}

\begin{figure}[]
\includegraphics[scale=1.1]{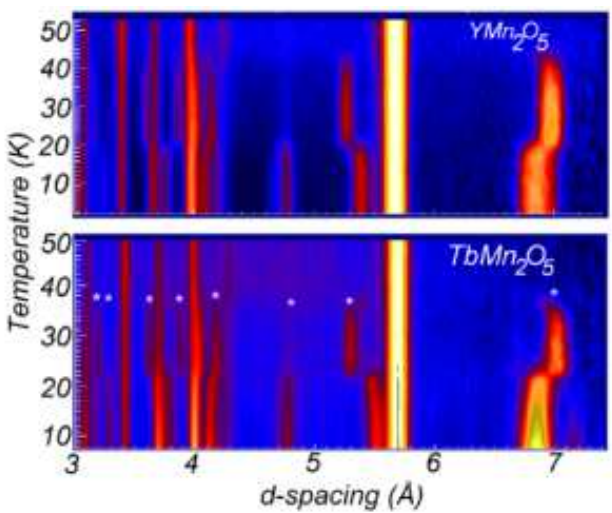}
\caption{Neutron powder diffraction patterns as a function of temperature for \Y (top) and \Tb (bottom) in the magnetically ordered phases. The scattering intensity is color coded with brighter colors representing higher intensities. The intensity of each diagram is normalized to the most intense nuclear reflections  in this d-spacing range ( (0,0,1) peak at 5.8 $\AA$). The position of magnetic Bragg peaks are indicated by white asterix.
}
\label{thermodiffractogram}
\end{figure}

The \rmno structure type has been synthesized with Y, all the lanthanides (excluding Ce) \cite{ISI:A1997YB24600016,ISI:000179358200010,ISI:000226555400064}, Bi \cite{ ISI:000174980300078} and La \cite{ISI:A1997YB24600016, ISI:000227257100009}.  Compounds containing lanthanum and all the rare earth lighter than Nd (included) do not become ferroelectric.
The "typical" features of the magnetic phase diagram of the \rmno family are best described based on the temperature-dependent neutron diffraction patterns in Fig. \ref{thermodiffractogram}. We can identify three or four regions depending whether the R site is magnetic or not. In the high-temperature paramagnetic phase, only nuclear Bragg peaks are present. On cooling, magnetic Bragg peaks first appear in reciprocal space positions corresponding to an incommensurate propagation vector $q=(\frac{1}{2}- \delta_x, 0, \frac{1}{4}+\delta_z)$. The values of $\delta_x$ (resp. $\delta_z$) range from 0.012 to 0.028 (resp. -0.013 to 0.027) \cite{ISI:000221010900040,ISI:000222268400032, ISI:000225913500038, ISI:000226362500075,ISI:000226555400064,ISI:000230276600039,ISI:000224662700076,ISI:000254542700065,ISI:000231564400011}. Within this high-temperature incommensurate phase(HT-ICP), the propagation vector changes with temperature, as it can clearly be seen from the curved shape of the Bragg peak centroid (this is best seen on the peak (0 1 0)$\pm$q around $ 7 \AA$).  On further cooling, the values of $\delta_x$ and $\delta_z$ smoothly approach zero, and the magnetic structure "locks" into a commensurate phase (CP).  The CP is the main ferroelectric phase  for most compositions.  Below $\sim 20 K$, there is an abrupt transition to a low-temperature incommensurate structure (LT-ICP) --- the position of the Bragg peaks continues to change down to the lowest temperatures.  Upon entering the LT-ICP, the value of the electrical polarization drops to a much smaller value \cite{ISI:000221644600033}, although there seems to be consensus that $P\neq 0$ in some cases (e.g., Y). For Tb, as well as other magnetic rare earths, the intensity of the magnetic peaks rapidly increased below $\sim 10 K$, indicating that the rare earth sublattice is becoming magnetically ordered (rare-earth ordered phase --- ROP).  The ROP is more strongly ferroelectric than the LT-ICP, and displays very large magneto-electric effects \cite{ISI:000248758000006,ISI:000221644600033,ISI:000223717600061}. This general magnetic phase diagram, with some subtleties, is shared by \rmno with $R$=Y, Ho,Tb and Er \cite{ISI:000230276600039,ISI:000225913500038,ISI:000221010900040,ISI:000248244500039,ISI:000224662700076,ISI:000235905700072}, which are the most studied compounds. Excluding the obvious absence of rare earth ordering for Y and the somewhat different transition temperatures, the most significant differences between these compounds are within the HT-ICS. For Ho and Tb, the two components $q_x$ and $q_z$ of the propagation vector seem to become commensurate at the same temperature, whereas for Er and Y there is a small region where $q_z=\frac{1}{4}$ while $q_x$ remains incommensurate \cite{ ISI:000248244500039}. The connection between this phenomenon and the appearance of "weak" ferroelectricity have not thoroughly been explored. In fact, hardly anything is know about the HT-ICS other than the propagation vectors, as the small magnetic moments make neutron studies more difficult than in the other phases. Three more compounds were studied in some detail:  Non-ferroelectric LaMn$_2$O$_5$ orders with a propagation vector $q=(0,0,\frac{1}{2})$ \cite{ISI:000227257100009,ISI:000254542700065}; DyMn$_2$O$_5$ behaves like the other systems with magnetic rare earths above 8.4 K, but displays commensurate ordering with $q=(\frac{1}{2},0,0)$ below this temperature, coexisting with a small fraction of the LT-ICP \cite{ISI:A1981LP35700027,ISI:000254542700065,ISI:000231564400011} ; the magnetic structure of BiMn$_2$O$_5$ is commensurate and ferroelectric at all temperatures with a propagation vector $q=(\frac{1}{2},0,\frac{1}{2})$ \cite{ ISI:000174980300078,vecchini}.

\section{Influence of an applied magnetic field}

\begin{figure}
\includegraphics[scale=1.4]{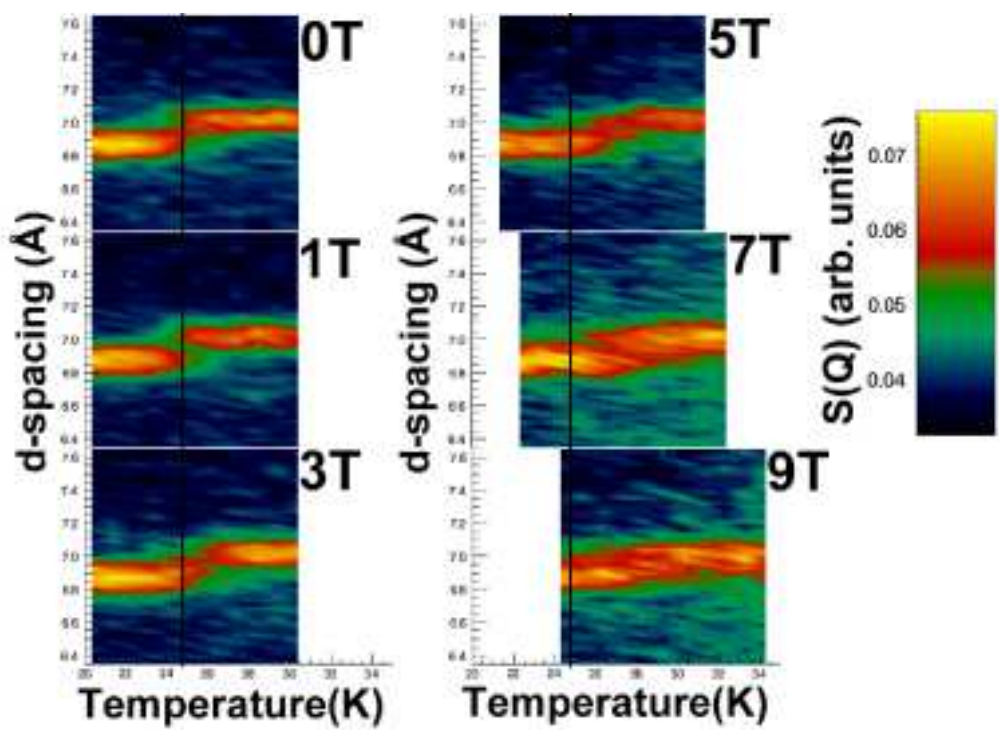}
\caption{Dependence of the (100)-k magnetic reflection with
temperature and applied magnetic field for TbMn$_2$O$_5$. The
scattering intensity is color coded and displayed as a function of
d-spacing and temperature for several values of the magnetic field.
The color scale is shown on the right. A vertical line is displayed
to mark the zero-field ICP-CP transition temperature (Color online)
}
\label{figurefield}
\end{figure}

\indent The strong magnetoelectric coupling in these materials can be directly evidenced by measurements of electric properties under magnetic field. For example in \Tb\, the upward jump in the dielectric constant at the CP to LT-ICP transition is pushed to higher temperatures as a magnetic field is applied \cite{ISI:000221644600033}, suggesting that the LT-ICP phase is stabilized by a magnetic field. This can be directly verified by constructing the H-T phase diagram of the CP-LT-ICP magnetic transition from powder neutron diffraction experiments in field. Data collected under fields up to 9 Tesla are reported in Fig. \ref{figurefield}. Zero-field data show that the
transition from CP to ICP phases is first order, as expected from group theory since the CP and ICP phases are characterized by wave-vectors belonging to different symmetry points of the Brillouin zone. The region of co-existence is very narrow
and can be estimated to be about 1-2K. Under application of a magnetic field, the CP-ICP transition temperature is enhanced from $\sim$ 25K in zero-field to more than 27K at 9 Tesla. The H-T phase
diagram, reported in Fig. 4, has been constructed by fitting the
data with two peaks of constrained widths, at positions fixed to the
CM and ICP peak positions in zero-field. The transition temperature
has been defined as the point of crossover between the intensity of
these peaks. The stabilization of the LT-ICP phase
under magnetic field is in perfect agreement with the magnetic and
dielectric phase diagrams \cite{ISI:000221644600033}, showing a gradual shift towards high
temperatures, on one hand, of the kink observed in the first
derivative of the magnetic susceptibility and, on the other hand, of
the upward jump in the dielectric constant. The CP to LT-ICP
transition is also smoothed out by magnetic field. Under field, the temperature
range of the CP/LT-ICP phases co-existence is extended, while the propagation
vectors of both phases do not seem to be altered. Overall, the
magnetic behavior obtained by powder neutron diffraction data
shows that the low temperature ICP phase is stabilized by
application of an external field. This result is opposite to
that recently published on the analog compound HoMn$_2$O$_5$
\cite{ISI:000242210200011}, showing that the boundary between CP and ICP phase
is shifted to lower temperature under application of a magnetic
field along the \textit{b}-axis. In both cases (Tb, Ho), neutron
data correlates directly to the electric properties under magnetic
field, confirming fundamentally different behaviors. Nevertheless,
it should be noted that experiments reported for the Ho system
\cite{ISI:000242210200011} and here have been conducted in
different conditions. Work on HoMn$_2$O$_5$ was performed on a
single crystal and only a magnetic field parallel to the
\textit{b}-axis was found to induce the ICP-CP transition. Work on powders is more qualitative due to the random
orientation of crystallites with respect to the magnetic field direction.
However, the general trend should not be affected by powder
averaging, since magnetic fields along \textit{a} and \textit{c}
proved to affect only slightly the magnetic state of \Tb\ . On the
other hand, for HoMn$_2$O$_5$, the induced CP state at high field is
observed only in field cooled experiments, whereas zero-field cooled
process leave the low-temperature ICP phase unchanged up to 13
Tesla. This indicates a large hysteresis (greater than 2 Tesla), as
pointed out by Kimura et al., due to the first order nature of the
transition. This does not explain, however, the opposite variation of
the CP-ICP temperature boundary in the H-T phase diagram, and it is more
likely that the discrepancy is a direct consequence of different
single-ion anisotropies of the rare-earth ion (R) and its related
effects on the Mn(d)- R(f) coupling. Another perspective in the comparison between Ho and Tb systems is
given by their magnetic behavior at low temperatures. Tb ions order magnetically below 10K with the
same propagation vector describing ordering of the Mn sublattice,
i.e. \textbf{k}$\sim$(0.48,0,0.31) \cite{ISI:000224662700076}. Under a moderately low magnetic
field (2.5 Tesla), Tb orders ferromagnetically, i.e at \textbf{k}=0,
without any noticeable change in the magnetic structure of the Mn
ions. This behavior is in sharp contrast with the results on Ho
\cite{ISI:000242210200011}, where it appears that the magnetic ordering of the
Mn and Ho sublattices arise at a single point in reciprocal space
(CP or ICP) irrespective of the value of the magnetic field,
suggesting a stronger d-f coupling. Moreover, the ferroelectric
properties of TbMn$_2$O$_5$, can be understood by the simple
superimposition of several order parameters associated with the Mn
and Tb sublattices respectively\cite{ISI:000221644600033}, implying the absence of
coupling terms between them. We suggested that magnetic
ordering of the Mn ions in \Tb\ is unaffected by the field-induced
FM ordering of Tb based on indirect evidence, since the
substraction of the 15 K zero-field data from the 1.5 K, 2.5 T data,
left an essentially perfect ferromagnetic pattern. This behavior can be verified by
in-field neutron diffraction experiment on the
analog YMn$_2$O$_5$. This compound shows the exact same sequence of
CP-ICP transitions on cooling, but containing a non-magnetic R site. Figure \ref{figurefield3}
presents powder neutron diffraction patterns obtained in zero-field
and in field up to 8 Tesla. The positions of the magnetic Bragg
peaks remains unchanged under magnetic field, indicating that the
propagation vector of the Mn magnetic structure does not vary. Also
the magnetic peak intensities do not change within the error bars,
confirming that the orientation of the magnetic moments remains unchanged. These results are in
agreement with magnetization measurements in very high magnetic
field, showing an extremely stable AFM structure at low temperature
and absence of metamagnetic transitions for the Mn sublattice up to
40 Tesla. Taken together with previous experiments on TbMn$_2$O$_5$,
this tends to favor a picture in which the Mn and Tb magnetic
sublattices are fully decoupled, since the metamagnetic transition
of Tb does not influences the magnetic ordering of Mn. The situation
for the Ho compound is therefore significantly different.

\section{Magnetic structures}

The magnetic structures of \rmno-type compounds have been studied since the seventies \cite{ISI:A1973P639000020,ISI:A1973P583900022}, and later in the eighties \cite{ISI:A1988R221000009} when the interest for them was mainly their unusual complexity.  A continuous improvement of experimental and analytical techniques has lead to a refinement in their understanding, which, in the case of the incommensurate phases, has not reached completion yet (see section \ref{section:conclusions}). The magnetic structure is best understood in terms of magnetic $a-b$ planes that are coupled along the $c$-axis.  All the ferroelectric phases share closely-related in-plane structures, characterized by the doubling (for the CP) or near-doubling (for the ICP) of the magnetic $a$ axis, while for non-ferroelectric LaMn$_2$O$_5$ (propagation vector $(0,0,\frac{1}{2})$  \cite{ISI:000227257100009} the $a$ axis is not doubled.  In particular, for all the commensurate, strongly ferroelectric phases, the in-plane magnetic structures are essentially identical.  This suggest that the in-plane arrangement of the spins plays a key role in inducing ferroelectricity.  On the contrary, the stacking of magnetic layers along the $c$ axis varies considerably from one compound to the next --- BiMn$_2$O$_5$ \cite{ISI:000174980300078,vecchini} and DyMn$_2$O$_5$ \cite{ISI:000230276600039,ISI:000254542700065,ISI:A1981LP35700027} have a simple $\cdot \cdot \cdot +-+-\cdot \cdot \cdot$ and $\cdot \cdot \cdot++++\cdot \cdot \cdot$ staking, respectively, whereas for the other phases the stacking is either exactly (CP) or approximately (ICP) $\cdot \cdot \cdot++--++--\cdot \cdot \cdot$.

As a typical example, the CP structure of YMn$_2$O$_5$ is shown in Fig \ref{magnetic} --- the figure also illustrated the symmetry-inequivalent magnetic interactions and their coupling constants.  In understanding how this structure is established, the following factors need to be taken into account:

\begin{enumerate}
\item{} \underline{Exchange interactions.}  Although no detailed calculation has been presented so far, the in-plane interactions are likely to be uniformly antiferromagnetic. The interactions along the $a$ direction are particularly strong, so that the presence of AFM zig-zag chains running along this direction is a common motif of all phases. For the LT-ICP, all magnetic structure determinations so far presented agree in evidencing a long-wavelength modulation of this motif, although the details of this modulation differ.
\item{} \underline{Frustration.}  The simultaneous presence of antiferromagnetic interactions and of \emph{fivefold} exchange rings in the crystal structure immediately leads to a frustrated situation, where not all these interactions can be satisfied simultaneously.  In the CP and ICP phases, this situation is relieved in different ways, possibly involving relaxation of the lattice and electronic structures and phasing of the zig-zag chains. Some of the resulting magnetic structure are acentric and capable of supporting ferroelectricity through magneto-elastic or magneto-electronic coupling.  Although the matter is still controversial, we believe that this relaxation provides the primary explanation for the ferroelectric phase diagram of these materials.
\item{} \underline{Anisotropy}  It is well-known that \Mt ions with low-lying $t_{2g}$ orbitals tend to be magnetically anisotropic. For octahedral coordination (e.g., in LaMnO$_3$, \cite{ISI:000072116400003}), the anisotropy is related to the Jahn-Teller distortion --- the spin direction \emph{parallel} to the unfilled $d_{z^2}$ orbitals is favored. In \Tbun and other cycloidal multiferroics, the interplay between anisotropy (favoring collinear magnetic structures) and competing interactions (favoring cycloidal magnetic structures) is thought to be primarily responsible for the magnetic phase diagram --- the former and latter arrangement are prevalent at high and low-temperature, respectively. In \rmno, \Mt has a pyramidal coordination, where the effect of the crystal field is very similar to the Jahn-Teller octahedral case.  Consequently, the \rmno compounds also display significant anisotropy:  in the CP, spin in each of the zig-zag chains are parallel to each other and to the axes of the pyramids. The axes of the pyramids are rotated by about $\pm 15 ^{\circ}$ away from the $a$ axis, explaining why the $a$ axis component of the spins, $S_x$ is much larger than the other two and also why the CP magnetic structure is slightly non-collinear.
\item{}\underline{$c$-axis stacking}.  As shown in Fig. \ref{nuclearstructure} and Fig. \ref{magnetic}, the stacking along the $c$ axis is mediated by crystallographic chains of edge-sharing Mn$^{4+}$O$_6$ octahedra.  To avoid confusion with the zig-zag chains running along the $a$ axis, we will refer to these structures as "$c$ axis ribbons".  Here, there are two relevant magnetic interactions:  through the \Mt layers ($J_2$) and through the $R$ layers ($J_1$) as shown in Fig. \ref{exchangealongchains}.  The \Mt - \Mf superexchange interactions through the common oxygen atoms appear to be the strongest contributors to $J_2$ which is therefore always ferromagnetic regardless of the sign of the in-plane interactions $J_3$ and $J_4$ (Fig. \ref{nuclearstructure}).  Consequently, in all phases, the spins in \Mt ions adjacent through a \Mf layer are \emph{always} close to be parallel. We take advantage of this to simplify the description of the $c-$ axis stacking, so that a single sign $+$ or $-$ actually represents two quasi-parallel \Mt sites.  The situation with the coupling $J_1$ through the $R$ layer is significantly more complex, explaining the diversity of the $c$-axis components $q_z$ of the magnetic propagation vectors.  In the case of \Bi and \Dy, $J_1$ is AFM or FM, respectively, and strong enough to enforce antiparallel ($\cdot \cdot \cdot +-+-\cdot \cdot \cdot$, propagation vector $q=(\frac{1}{2},0,\frac{1}{2})$) or parallel ($\cdot \cdot \cdot++++\cdot \cdot \cdot$, propagation vector $q=(\frac{1}{2},0,0)$) staking, respectively.  In all the other cases, $J_1$ must be small, since both quasi-parallel and quasi-antiparallel stackings are found within the same magnetic structures.
\item{}\underline{NNN and antisymmetric interactions}  At present, it is not known what determines the $c-$ axis stacking, and the corresponding $q_z$ in materials other than \Dy and \Bi. Two main effects are thought to be responsible for complex stacking and incommensurability in magnetic materials : competition between nearest- and next-nearest-neighbor interaction and the antisymmetric \DM  (DM) coupling. The former is the accepted mechanism to explain the cycloidal phases in \Tbun and other cycloidal multiferroics. An analogous model that can be applied to \R\ will be discussed in section \ref{section: NN_NNN}. It has recently been shown \cite{ISI:000248244500035,vecchini} that in the CP of several \rmno compounds, including \Bi, a small $c$-axis component $S_z$ of the spins is present, and that in the first approximation this is out-of-phase with respect to the other two components, giving rise to very flat cycloids running along the $c$ axis.  It is interesting to remark that this component is present \emph{regardless} of the propagation vector, although it is smaller in the case of \Bi.  For \Bi, there is a very simple explanation for the appearance of a $c$-axis component of this kind, due to the antisymmetric DM interaction (see Section \ref{section: DM} below).  It is tempting to extend this explanation to the other phases as well, and conclude that CP flat cycloids are of DM origin.  We remark, however, that no detailed models or calculations have been presented so far.
\end{enumerate}

\indent Leaving the small $S_z$ component aside, the general features of the in-plane magnetic structures of \rmno in the CP regime can be rationalized very well based on points (i) - (iii) above.  From Fig. \ref{nuclearstructure} and Fig. \ref{magnetic}, it can clearly be seen that in the absence of structural distortions, there is an exact cancelation of all energy terms containing $J_3$, since pairs of ions related by inversion symmetry carry magnetic configurations of opposite sign. This situation gives rise to the so-called magnetic Jahn-Teller effect, due to the analogy with the well-known structural effect. Here, the system can always gain energy from distortion of the crystal or electronic structure that makes the $J_3$ slightly inequivalent, since the energy gain is linear and the energy cost is typically quadratic in the distortion.  We have previously argued \cite{ ISI:000224662700076,ISI:000235905700072} that this is the primary mechanism leading to the appearance for ferroelectricity in \rmno (see next section).

\section{Origin of ferroelectricity}
\label{section:origin}
Two main models have been proposed to explain the appearance of ferroelectricity in the CP of \rmno; here, we will refer to these models as the symmetric exchange-striction model and the cycloidal model, respectively.  These two models are at present very difficult to disentangle based on the facts know to us with confidence.  In particular, the point-group symmetry of the CP, which has been solved with great accuracy, is still very high ($m2m$, see Ref \cite{ISI:000249155100099}), so the ferroelectric polarization \emph{must} lie along the $b$ axis, regardless of the mechanism.

The exchange-striction model relies on the magnetic Jahn-Teller effect as its main ingredient: here, electrical polarization would arise from a combination of atomic displacements and electronic rearrangements, removing the exact exchange degeneracy. The main features of the exchange-striction model are:

\begin{itemize}
\item{} The most important element controlling ferroelectricity is the in-plane components of the magnetic structure.  Non-collinearity is not an essential ingredient, as it has been shown that the same effect can be obtained in a collinear acentric structure. The stacking along the $c$ axis is immaterial and so is the small $c$ axis component, provided that all the layers have the same polarization.
\item{} Quantitatively, the electrical polarization is proportional to the \emph{scalar} product of spins related by $J_3$ in different zig-zag chains.  Therefore, the relative phasing of the two chains is directly related to the magnitude of the polarization \cite{ISI:000235905700072}. In the CP, this relative phase is uniform, so that each $J_3$ pair contributes equally to ferroelectricity. However, there is no reason for this to be so in the LT-ICP --- in fact for all models of the ICP so far proposed the \emph{scalar} product varies in both sign and magnitude along the propagation direction. This provides an immediate explanation to the sudden loss of ferroelectricity in the LT-ICP (see below).
\item{} As mentioned before, in the CP the electrical polarization is \emph{always} directed along the $b$ axis by symmetry. With all probability, there is a similar requirement for the incommensurate phases, so that whatever residual polarization remains in the LT-ICP should also be directed along the $b$ axis, in agreement with all the experiments. Very recent electronic structure calculations have provided \cite{ ISI:000250506000061,wang2} evidence that ferroelecticity can be explained without invoking the spin-orbit coupling.
\end{itemize}

The cycloidal model relies on the same spin-orbit-driven, \emph{inverse} \DM effect that has been used to explain the physics of \Tbun and other cycloidal magnets:  here, electrical polarization would arise again from a combination of atomic displacements and electronic rearrangements, but this time what is minimized is the \emph{antisymmetric} interaction between non-collinear spins. The general expression for the polarization thus generated is $\textbf{P} \propto \textbf{e}_{12} \times ( \textbf{S}_1 \times \textbf{S}_2)$, where $\textbf{S}_1$ and  $\textbf{S}_2$ are spins on adjacent sites along the $c$ axis and $\textbf{e}_{12}$ is a unit vector connecting them (see for example \cite{ISI:000254144700013}).

The main features of the cycloidal model are:

\begin{itemize}
\item{} It provides a unified, "universal" explanation for all novel multiferroics. This is perhaps the most important reason of its popularity.
\item{} Non-collinearity in the context of a cycloidal structure is an essential ingredient of the model.
\item{} Locally, the polarization direction is defined by the cross product of two vectors, \textbf{u}, which is perpendicular to the plane of rotation of the spins, and \textbf{q}', defined as the projection of the propagation vector \textbf{q} along the direction of the bonds contributing to antisymmetric exchange.  \textbf{q}' may be difficult to define in general, because several bonds could provide independent contributions, but in the case of the CP, \textbf{q}' is directed along the $c$ axis. The contribution to the electrical polarization of different ribbons may cancel in some directions. For example, in the CP, the plane of rotation of the flattened cycloids is defined by the $c$-axis and by the anisotropy direction, which is rotated $15^{\circ}$ away from the $a$ axis.  However, the large $P_x$ components cancel out between the two ribbons in the unit cell, whereas the smaller $P_y$ component add, yielding an overall polarization along the $b$ axis. Once again, this is entirely due to symmetry.
\item{} Preliminary investigations\cite{vecchini_ICM} indicate that the LT-ICP has lower symmetry than the CP, but the polarization is still required to lie along the $b$ direction.  A LT-ICP precise solution of the magnetic structure may be able to discriminate conclusively between the two mechanisms.
\end{itemize}

Most of the consideration made so far are qualitative. However, it is possible to make a quantitative assessment of the likely contributions of both exchange-striction and cycloidal mechanisms, once the magnetic structures are known in detail. What can never be known from the magnetic structure alone are the coupling constants for symmetric and antisymmetric exchange, although it is reasonable to assume that the antisymmetric coupling should be weaker, since it relies on the intrinsically weak spin-orbit mechanism. In both cases, the electrical polarization \textbf{P} is obtained by multiplying the appropriate coupling constant times a polar vector (\textbf{E} or \textbf{S} below), which is entirely defined by the magnetic structure and has the dimension of $\mu_B^2$. For the calculation of these polar vectors, only the spins within one unit cell need to be taken into account, once the propagation vector is known. Therefore, the same formalism can be applied to the CP and to all the ICPs.
For the calculations presented in this section, we employ the following convention: the three components of the magnetic moment of a given site s, V$_i^s$ (i=x,y,z) in unit cell R$_l$ are:
\begin{equation}
V_i^s(R_l)=M_i^s\,cos(qR_l+\phi_i^s)
\end{equation}
The M$_i$s and $\phi_i$s can be calculated from the Fourier coefficient given in \cite{vecchini}. We note that the propagation vector q and phase $\phi$ are given here in radians and not in fractional units of 2$\pi$ \cite{vecchini}. The same labels than \cite{ISI:000230276600039,vecchini} are employed for magnetic ions in the unit-cell.\\
\indent The exchange striction polar vector $E_y$ is always directed along the $b$ axis.  If $q_x$ is incommensurate,  $E_y$ is calculated as

\begin{eqnarray}
\label{eq: ES_ICP}
E_y&=&\frac{1}{2} \,\sum_{i=1,3} M_i^{b2}\,M_i^{a1}\,\cos(\phi_i^{b2}-\phi_i^{a1})+M_i^{b1}\,M_i^{a1}\,\cos(\phi_i^{b1}-\phi_i^{a1})\nonumber\nonumber\\
&&-M_i^{b2}\,M_i^{a4}\,\cos(\phi_i^{b2}-\phi_i^{a4}+q_x)-M_i^{b1}\,M_i^{a4}\,\cos(\phi_i^{b1}-\phi_i^{a4}+q_x)\nonumber\\
&&+M_i^{b3}\,M_i^{a2}\,\cos(\phi_i^{b3}-\phi_i^{a2})+M_i^{b4}\,M_i^{a2}\,\cos(\phi_i^{b4}-\phi_i^{a2})\nonumber\\
&&-M_i^{b3}\,M_i^{a3}\,\cos(\phi_i^{b3}-\phi_i^{a3})-M_i^{b4}\,M_i^{a3}\,\cos(\phi_i^{b4}-\phi_i^{a3})
\end{eqnarray}

If $q_x=\pi$, there is an additional "umklapp" term, so that

\begin{eqnarray}
\label{eq: ES_CP}
E_y&=&\sum_{i=1,3} M_i^{b2}\,M_i^{a1}\,\cos\phi_i^{b2}\cos \phi_i^{a1}+M_i^{b1}\,M_i^{a1}\,\cos\phi_i^{b1}\cos\phi_i^{a1}\nonumber\nonumber\\
&&-M_i^{b2}\,M_i^{a4}\,\cos\phi_i^{b2}\cos(\phi_i^{a4}+q_x)-M_i^{b1}\,M_i^{a4}\,\cos\phi_i^{b1}\cos(\phi_i^{a4}+q_x)\nonumber\\
&&+M_i^{b3}\,M_i^{a2}\,cos\phi_i^{b3}\cos\phi_i^{a2}+M_i^{b4}\,M_i^{a2}\,\cos\phi_i^{b4}\cos\phi_i^{a2}\nonumber\\
&&-M_i^{b3}\,M_i^{a3}\,\cos\phi_i^{b3}\cos\phi_i^{a3}-M_i^{b4}\,M_i^{a3}\,\cos\phi_i^{b4}\cos\phi_i^{a3}
\end{eqnarray}

There are two separate components of the spin-orbit polar vector, referring to the cross product of spins across the Mn layer ($S^1$) or across the \textit{RE} layer ($S^2$).  The same formula serves to calculate both $x$ and $y$ components.  For $q_z \neq \pi$ we obtain:

\begin{eqnarray}
\label{eq: SO_ICP}
S^1_i&=&\frac{1}{2}\left[M^{b1}_i\,M^{b_2}_z\,\cos(\phi^{b1}_i-\phi^{b2}_z)-M^{b1}_z\,M^{b2}_i\,\cos(\phi^{b1}_z-\phi^{b2}_i)\right.\nonumber\\
&&\left.+M^{b3}_i\,M^{b4}_z\,\cos(\phi^{b3}_i-\phi^{b4}_z)-M^{b3}_z\,M^{b4}_i\,\cos(\phi^{b3}_z-\phi^{b2}_i)\right.\nonumber\\
S^2_i&=&\frac{1}{2}\left[M^{b2}_i\,M^{b_1}_z\,\cos(\phi^{b2}_i-\phi^{b1}_z-q_z)-M^{b2}_z\,M^{b1}_i\,\cos(\phi^{b2}_z-\phi^{b1}_i-q_z)\right.\nonumber\\
&&\left.+M^{b4}_i\,M^{b3}_z\,\cos(\phi^{b4}_i-\phi^{b3}_z-q_z)-M^{b4}_z\,M^{b3}_i\,\cos(\phi^{b4}_z-\phi^{b3}_i-q_z)\right]
\end{eqnarray}

Whereas for the umklapp case $q_z = \pi$ of \Bi  we obtain:

\begin{eqnarray}
\label{eq: SO_CP}
S^1_i&=&M^{b1}_i\,M^{b_2}_z\,\cos\phi^{b1}_i\cos\phi^{b2}_z-M^{b1}_z\,M^{b2}_i\,\cos\phi^{b1}_z\cos\phi^{b2}_i\nonumber\\
&&+M^{b3}_i\,M^{b4}_z\,\cos\phi^{b3}_i\cos\phi^{b4}_z-M^{b3}_z\,M^{b4}_i\,\cos\phi^{b3}_z\cos\phi^{b2}_i\nonumber\\
S^2_i&=S^1_i
\end{eqnarray}

The total polarization is given by

\begin{eqnarray}
P_x&=&c^{so}_1\,S^1_x+c^{so}_2\,S^2_x\nonumber\\
P_y&=&c^{es}\,E_y+c^{so}_1\,S^1_y+c^{so}_2\,S^2_y
\end{eqnarray}

where $c^{so}_1$, $c^{so}_2$ and $c^{es}$ are magneto-elastic coupling constants and can have either sign.

\begin{table}
\begin{tabular}{ccccccc}
  Compound & $\textbf{q}$ &$E_y$ & $S^1_x$ & $S^1_y$ & $S^2_x$ & $S^2_y$ \\
\hline\\
\Y CP \cite{vecchini} & $(\frac{1}{2},0,\frac{1}{4})$ &22.6 & -0.01 & -0.16 & -0.04 & -0.39\\
\Y ICP \cite{Kim} & $(0.48,0,0.29)$ & 6.8 & -5.84 & 1.34 & -4.88 & -0.49\\
\Ho CP \cite{vecchini} & $(\frac{1}{2},0,\frac{1}{4})$ &27.4 & -0.10 & -0.23 & -0.21 & -0.46\\
\Bi CP \cite{vecchini}& $(\frac{1}{2},0,\frac{1}{2})$ &-45.4 & -1.38 & 0.08 & -1.38 & 0.08\\
\end{tabular}
\caption{Magnetic polar vector components relevant for the exchange-striction mechanism ($E_y$) and the spin-orbit mechanism ($S^1_x$,  $S^1_y$ , $S^2_x$ and $S^2_y$), calculated from Eq. \ref{eq: ES_ICP}-\ref{eq: SO_CP} for different \rmno compounds.  We employed our data for the CPs, whereas for the ICP of \Y we used recent data by Kim \textit{et al.} \cite{Kim}.  All values are in $\mu_B^2$.  The different signs of $E_y$ are not significant, since they are due to the inversion-domain variant chosen for the refinement.
}
\label{tab:mechanisms}
\end{table}

Calculated data for the different polar vector components based on published magnetic structures are reported in Tab. \ref{tab:mechanisms}.  The following observations can be made:

\begin{enumerate}
\item{}In the strongly ferroelectric CPs, the exchange-striction polar vector $E_y$ is always much larger than the cycloidal polar vectors --- typically by a factor of 50-100. This is in itself a strong indication that exchange striction should be the dominant mechanism if one considers the fact that the exchange-striction coupling constants are also expected to be larger.
\item{} We have reported for comparison the polar vector components for the most recently published refinement of the \Y ICP by Kim \textit{et al.} \cite{Kim}. This structure contains cycloids in the $a-b$ plane as well, which could in principle contribute to the ferroelectricity through the spin-orbit coupling mechanism.  We observe, however, that the strong decrease of $E_y$ is in accord with the drop in the electrical polarization, as observed experimentally. The decrease of $E_y$ is due to the fact that the zig-zag chains are now out-of-phase rather than in-phase, as we remarked for the ICP structure we previously obtained from powder data \cite{ISI:000235905700072}. On the contrary, the trend on the cycloidal polar vector components --- in particular the prediction of a significant polarization along the $a$ axis --- is inconsistent with the experiments.
\item{} The large value of $E_y$ calculated for \Bi is in agreement with the experimentally observed large polarization. However, our magnetic structure would be consistent with a spin-orbit polarization directed along the $x$ axis, at odds with the experiments. Nevertheless, a more precise determination of the $s_z$ components, which are very small for \Bi, would be required to be able to employ this argument with confidence to validate the exchange-striction model.

\end{enumerate}

\section{A simple model for the propagation along the $c$ axis}
\label{section: NN_NNN}

The magnetic phase diagram of the \rmno compounds has been studied phenomenologically in great detail by Harris and coworkers \cite{harris_125_1,harris_125_2}. In this section, we propose a more modest approach, based on the well-known linear-chain model, which, however, contain the minimal ingredients --- competition between nearest-neighbor (NN) and next-to-nearest neighbor (NNN) interactions, widely thought to be responsible for the observed cycloidal structures in \Tbun and related compounds. The purpose of this discussion is to ascertain whether these ingredients are sufficient to explain the general tendency of \rmno to develop an out-of-phase $s_z$ component for a variety of propagation vectors.

The spins and interactions to be considered , as  shown in Fig \ref{exchangealongchains}, are:

\begin{eqnarray} \label{eq: spins}
\textbf{s}^{\prime}_3&=& \hat{\textbf{i}} \, s_{x}\,\cos\left(\beta \, R_z - \alpha/2 + \beta \right) + \hat{\textbf{k}}\, s_{z}\,\sin\left(\beta  \, R_z - \alpha/2 + \beta \right)\nonumber\\
\textbf{s}_1&=& \hat{\textbf{i}} \, s_{x}\,\cos\left(\beta  \, R_z + \alpha/2 \right) + \hat{\textbf{k}}\, s_{z}\,\sin\left(\beta  \, R_z + \alpha/2 \right)\nonumber\\
\textbf{s}_2&=& \hat{\textbf{i}} \, s_{x}\,\cos\left(\beta  \, R_z - \alpha/2 \right) + \hat{\textbf{k}}\, s_{z}\,\sin\left(\beta  \, R_z - \alpha/2 \right)\nonumber\\
\textbf{s}_3&=& \hat{\textbf{i}} \, s_{x}\,\cos\left(\beta  \, R_z + \alpha/2 - \beta \right) + \hat{\textbf{k}}\, s_{z}\,\sin\left(\beta  \, R_z + \alpha/2 - \beta \right)
\end{eqnarray}

where $\hat{\textbf{i}}$ and $\hat{\textbf{k}}$ are unit vector in the $xy$-plane and $z$ directions, respectively, $R_z$ is the unit cell index, $\beta $ is the propagation vector component along $z$ and $\alpha/2$ is the phase of the spins, the latter two being expressed in radians, so that they relate to the tabulated values as $\alpha= 4 \pi \delta$ and $\beta=2 \pi q_z$ \cite{vecchini}.  Here, $s_x$ and $s_z$ refer to a generic in-plane component and to an out-of-plane component, respectively.

We can construct an idealized 2-dimensional model, coupled  through NN interactions ($J_1$ and $J_2$, which are different) and NNN interactions ($J^{\prime}$).  The total energy per unit cell is

\begin{eqnarray} \label{eq: energy1}
E&=&\frac{1}{N}\sum_{R_z}{J_2 \, \textbf{s}_1 \cdot \textbf{s}_2 + \frac{1}{2}J_1 \, \left( \textbf{s}_1 \cdot \textbf{s}^{\prime}_3+ \textbf{s}_2 \cdot \textbf{s}_3\right) + \frac{1}{2}J^{\prime} \left( \textbf{s}_1 \cdot \textbf{s}_3 + \textbf{s}_2 \cdot \textbf{s}^{\prime}_3\right)}\nonumber\\
&&+ \Gamma \, s_z^2\,\left(\sin ^2\left(\beta  \, R_z + \alpha/2 \right)+ \sin ^2\left(\beta  \, R_z - \alpha/2 \right)\right)
\end{eqnarray}

The last term containing $\Gamma$ represents the contribution of the anisotropy.  By replacing the expressions of the spins from eq. \ref{eq: spins} and decomposing the energy into the normal and umklapp  terms $E=E_N+E_U$ we obtain:

\begin{eqnarray}\label{eq: energy_dec}
E_N&=&\frac{s_x^2+s_z^2}{2}\left(J_2 \cos \alpha + J_1 \cos (\alpha-\beta )+J^{\prime} \cos \beta \right)+\Gamma s_z^2\\
E_U&=& \frac{s_x^2-s_z^2}{2N} \sum_{R_z} \cos (2\beta R_z) \left(J_2 +J_1 \cos \beta +J^{\prime} \cos \left(\alpha - \beta \right) \right)\nonumber\\
&&-\Gamma \frac{s_z^2}{N} \sum_{R_z} \cos (2\beta R_z)\cos \alpha
\end{eqnarray}

One can easily see from eq. \ref{eq: energy_dec} that the umklapp term is non-zero only for $\beta = 0$ and $\beta = \pi$ (doubling of the unit cell along the $z$ axis).

Let us first consider the general case in which the umklapp term is zero.  We note that we can ignore the anisotropy when we minimize the energy as a function of the two angles $\alpha$ and $\beta$, since the anisotropy term does not contain the angles.  By defining $R=J^{\prime}/J_2$ and $T=J_1/J_2$ and setting the first derivatives to zero we obtain the two solutions:

\begin{eqnarray} \label{eq: solutions}
\sin \alpha &=& -R \, \sin \beta\nonumber\\
with \nonumber\\
\sin \beta&=&0\nonumber\\
or \nonumber\\
\cos \beta &=& \frac{T}{2} \left(\frac{1}{R^2}-1\right)-\frac{1}{2T}
\end{eqnarray}

The first solution \emph{always} gives rise to umklapp since $\beta = 0 $ or $ \pi$, so it should always be discarded.  We can always set $\beta > 0$, whereas $\alpha$ is set in the appropriate quadrant to satisfy eq. \ref{eq: solutions}.  Once this is done, one should obtain the values of $s_x$ and $s_z$ under the constraint that the total spin at any site cannot exceed the full ionic value ($3/2 \mu_B$ for \Mf).  For an elliptical cycloid of this type, the maximum spin value is $\max (s_x, s_z)=s_x$, i.e., the long semi-axis of the ellipse.  Therefore, regardless of the specific values of $\alpha$ and $\beta$, the minimum energy is always attained for either $s_z=s_x$ (small anisotropy, circular cycloid) or $s_z=0$ (large anisotropy, collinear spin density wave).

The two situations riving rise to umklapp are $\beta = 0$ and $\beta = \pi$.  We note that in this simple model we cannot obtain the $\beta = \pi/2$ lock-in situation that characterizes the ferroelectric phase.  In order for this to occur, one would need a quartic term in the free energy --- for example, through magneto-elastic interaction. The presence of these quartic terms manifests through the appearance of charge peaks at twice the magnetic propagation vector\cite{Beutier}.  Setting $\rho=s_z/s_x$, the two expressions to minimize in the presence of umklapp become:

\begin{eqnarray} \label{eq: umklapp_zero}
E&=&\frac{s_x^2}{2} \, \left [(J_1+J_2)(1- \rho^2) + J^{\prime}\,(1+ \rho^2)+2 \Gamma \rho^2\right.\nonumber\\
&&\left.+\cos \alpha \, \left[(J_1+J_2)(1+ \rho^2)+J^{\prime} \,(1-\rho^2)- 2\Gamma \rho^2\right] \right] \nonumber\\
&&\textrm{for} \,\, \beta = 0
\end{eqnarray}

and

\begin{eqnarray}\label{eq: umklapp_pi}
E&=&\frac{s_x^2}{2} \, \left [(-J_1+J_2)(1- \rho^2) - J^{\prime}\,(1+ \rho^2)+2 \Gamma \rho^2\right.\nonumber\\
&&\left.+\cos \alpha \, \left[(-J_1+J_2)(1+ \rho^2)-J^{\prime} \,(1-\rho^2)- 2\Gamma \rho^2\right] \right] \nonumber\\
&&\textrm{for} \,\, \beta = \pi
\end{eqnarray}

With the help of eq. \ref{eq: energy_dec},\ref{eq: solutions}, \ref{eq: umklapp_zero} and \ref{eq: umklapp_pi}  one can construct phase diagrams for what is effectively a two-parameter problem.  The case relevant for \rmno is that of $J_2 <0$, since, as we mentioned previously, the interaction through the \Mt layer is always ferromagnetic.  Fig. \ref{Fig:phase_diagram} shows the phase diagram at fixed $J_2 =-1$ as a function of $J_1$ and $J^{\prime}$ and in the isotropic case ($s_x= \pm s_z \, , \Gamma=0$).  We have plotted only the region in which $J_2$ is the largest exchange constant --- a plausible assumption, as we have seen, but the extended phase diagram is equally easy to plot.

The following observations can be made by inspecting the phase diagrams:

\begin{enumerate}
\item{}  \label{item:_1} Even in the absence of umklapp, there are two large regions of commensurability. For $\beta=0$ the magnetic cell coincides with the chemical cell, whereas for $\beta=\pi$ the magnetic cell is doubled.
\item{}\label{item:_2} More significantly, $\alpha=0$ \emph{always} whenever $\beta=0$ or $\beta=\pi$.  This is also clear by inspecting eq. \ref{eq: umklapp_zero} and \ref{eq: umklapp_pi}, since this expression is minimized for one of the extrema of $\cos \alpha$.  Since $\alpha=0$ in this part of the phase diagram, the solution is a collinear AFM, and the terms containing $\rho$ and $\Gamma$ in eq. \ref{eq: umklapp_zero} and \ref{eq: umklapp_pi} cancel out.  This is very important, since it indicates that this simple model is incapable of reproducing the \Bi situation, where the presence of a $s_z$ component indicates that $\alpha>0$ even with $\beta=\pi$ (propagation vector $\textbf{q}=(\frac{1}{2}, 0, \frac{1}{2})$.
\item{} \label{item:_3} As soon as one departs from the isotropic situation, the umklapp terms tend to stabilize the commensurate phases, and the region of incommensurability shrinks.  For $\rho=0.5$, the portion of the phase diagram shown in Fig. \ref{Fig:phase_diagram} is completely commensurate with $\alpha=0$, and is split diagonally (top left to bottom right) between $\beta=0$ (bottom) and $\beta=\pi$ (top).  Typical experimental values of $\rho$ are even smaller --- of the order of $0.2$.
\item{} \label{item:_4} Even in the presence of umklapp, one can always find small patches of incommensurability in the extended phase diagram.  However, in order to stabilize the \rmno incommensurate phase by this mechanism, one would require extremely fine-tuning of the exchange parameters.
\end{enumerate}

One also observes that there is complete degeneracy between the configurations differing by the sign of the $z$ components of the spins ($\alpha \rightarrow -\alpha\, , \, \beta \rightarrow - \beta$), corresponding to the counter-clockwise and clockwise rotation of the cycloids (if one imagines to travel along the positive $z$ direction, and to the two different polarities of the magnetic structure (there is no true chirality here).  This is completely trivial in this simple 1D model, but it is not so in the realistic situation of \rmno, since, as we have seen, the in-plane magnetic structure also has a polarity, and reversing all the $z$ components would correspond to a non-symmetry equivalent domain (see Section \ref{section:cryopad} below).  The degeneracy between these domains must be lifted to some degree, in order to enable the observation of a $s_z$ component by diffraction.

This observation, together with those at point \ref{item:_2} and \ref{item:_4} above, suggest that an additional mechanism, in addition to competition between NN and NNN interactions, may be required to explain the observed phase diagram of \rmno.

\section{\DM interaction and its consequences for the \rmno magnetic structures}
\label{section: DM}

As we have seen, the simple model presented in Section \ref{section: NN_NNN} is incapable of reproducing some important features found in \rmno compounds --- for example the presence of a small $s_z$ component in \Bi.  Moreover, opposite cycloidal polarities (i.e., opposite signs of $s_z$, everything else being equal) are energetically degenerate, while this degeneracy must be lifted to be consistent with the diffraction observations.  In fact, this degeneracy is characteristic of all models that contain symmetric exchange terms only, at least for phases that retain part of the crystallographic symmetry.  In fact, symmetric coupling energy terms that split the degeneracy must be of the form $S_x^iS_z^j$ or $S_y^iS_z^j$, where $i$ and $j$ refer to different magnetic sites. If a set of mirror planes perpendicular to the $c$ axis is retained, as it is the case, for example, for the commensurate phase of \rmno \cite{vecchini}, terms of this form cancel out exactly by symmetry. It is therefore useful to look specifically for an antisymmetric interaction term that is capable of lifting the degeneracy.

In the crystal structure of the paramagnetic phases, the DM vector $\textbf{D}$ between \Mf atoms chained along the $z$ axis is identically zero, because these sites are related by inversion (Fig. \ref{figureDM}). In the ferroelectric phase, this is no longer strictly true, because the ferroelectric displacements break inversion symmetry. However, these displacements are extremely small, and so should also be the resulting $\textbf{D}$ vector. Here we show that there is a much more efficient mechanism to generate non-collinearity along the $z$ axis, through an effective antisymmetric interaction mediated by the \Mt atoms. \Mt and \Mf atoms are not related by any symmetry, and the $\textbf{D}$ vector associated with pairs of such atoms is in no way restricted. However, $\textbf{D}$ transforms like an axial vector between symmetry-related bonds, so some of the DM interaction terms cancel out by symmetry. The labeling scheme we employ to calculate the DM energy is shown in Fig. \ref{figureDM}, and is consistent with \cite{vecchini}.

\begin{eqnarray} \label{eq: DM_symmetry}
\textbf{D}_{1-4}^{\parallel}&=&-\textbf{D}_{2-4}^{\parallel}=-\textbf{D}_{1-1}^{\parallel}=\textbf{D}_{2-1}^{\parallel}\nonumber\\
\textbf{D}_{1-4}^{\perp}&=&\textbf{D}_{2-4}^{\perp}=\textbf{D}_{1-1}^{\perp}=\textbf{D}_{2-1}^{\perp}\nonumber\\
\textbf{D}_{1-2}^{\parallel}&=&-\textbf{D}_{2-2}^{\parallel}=-\textbf{D}_{1-3}^{\parallel}=\textbf{D}_{2-3}^{\parallel}\nonumber\\
\textbf{D}_{1-2}^{\perp}&=&\textbf{D}_{2-2}^{\perp}=\textbf{D}_{1-3}^{\perp}=\textbf{D}_{2-3}^{\perp}
\end{eqnarray}

where $\textbf{D}^{\parallel}$ and $\textbf{D}^{\perp}$ are the components of the DM vector parallel/perpendicular to the $ab$ plane.  By employing Eq. \ref{eq: DM_symmetry} we can calculate the DM energy E$_{DM}$:

\begin{eqnarray} \label{eq: DM_energy}
E_{DM}&=&\textbf{D}_{1-4}^{\parallel} \cdot \left[\left(\textbf{S}_{a4}-\textbf{S}_{a1}\right) \times\left(\textbf{S}_{b1}-\textbf{S}_{b2}\right)\right]+\nonumber\\
&&\textbf{D}_{1-4}^{\perp} \cdot \left[\left(\textbf{S}_{a4}+\textbf{S}_{a1}\right) \times\left(\textbf{S}_{b1}+\textbf{S}_{b2}\right)\right]+\nonumber\\
&&\textbf{D}_{1-2}^{\parallel} \cdot \left[\left(\textbf{S}_{a2}-\textbf{S}_{a3}\right) \times\left(\textbf{S}_{b1}-\textbf{S}_{b2}\right)\right]+\nonumber\\
&&\textbf{D}_{1-2}^{\perp} \cdot \left[\left(\textbf{S}_{a2}+\textbf{S}_{a3}\right) \times\left(\textbf{S}_{b1}+\textbf{S}_{b2}\right)\right] = \nonumber\\
&\simeq&2 \,\textbf{D}_{1-4}^{\parallel} \cdot \left[\textbf{S}_{a4} \times\left(\textbf{S}_{b1}-\textbf{S}_{b2}\right)\right]+2 \, \textbf{D}_{1-2}^{\perp} \cdot \left[\textbf{S}_{a2} \times\left(\textbf{S}_{b1}+\textbf{S}_{b2}\right)\right]
\end{eqnarray}

In the last line of Eq. \ref{eq: DM_energy} we have exploited the fact that $\textbf{S}_{a4} \simeq -\textbf{S}_{a1}$ and $\textbf{S}_{a2}\simeq\textbf{S}_{a3}$ (the equality is exact for fully constrained models and $q_x=1/2$).
One should also remark that for the commensurate phases $\textbf{S}_{a2}$ is essentially in-plane --- it is in fact exactly so by symmetry for \Bi and approximately for most of the other elements with $\textbf{q}=(\frac{1}{2}, 0, \frac{1}{4})$. Therefore, the effect of the first term is clearly that of inducing a $z$ component of opposite sign, and therefore a canting angle between $\textbf{S}_{b1}$ and $\textbf{S}_{b2}$. It is an effective antisymmetric interaction along the chain, mediated by the in-plane spins. The second term induces an in-plane canting angle $\textbf{S}_{a}$ and $\textbf{S}_{b}$, spins. Since this does not break the symmetry, we will not consider it in the remainder.

We can reproduce the effect of the first term in our simple 1-dimensional model by making a correspondence between spins in Fig. \ref{exchangealongchains} and Fig. \ref{figureDM} and setting the DM vector along the $b$ axis, i.e., $\textbf{D}_{1-4}^{\parallel}=\hat{\textbf{j}}D$. We also assume $\textbf{S}_{a4}=\tilde{s}_x \, \cos \beta R_z $. After some manipulations we obtain:

\begin{eqnarray} \label{eq: DM_energy_model}
E_{DM}&=&\frac{2\,D}{N}\sum_{R_z}{\tilde{s}_x s_z\, \cos \beta R_z \left[\sin \left(\beta R_z-\alpha/2\right)-\sin\left(\beta R_z+\alpha/2\right)\right]}=\nonumber\\
&=& -D \tilde{s}_x s_z\, \sin \alpha/2
\end{eqnarray}

It is clear that $E_{DM}$ stabilizes a non-zero value of $\alpha$ even for the umklapp cases (Eq. \ref{eq: umklapp_zero} and \ref{eq: umklapp_pi}). This simple mechanism is therefore capable of explaining the observation of a small $s_z$ component in the case of \Bi, where $\beta=\pi$. We can see now that $E_{DM}$ discriminates between right-handed and left-handed cycloids, since its sign is reversed if one changes the sign of $\alpha$. In addition, the sign of $\alpha $ (reversal of cycloidal polarity) must change in response to a change in the sign of $\tilde{s}_x$ . For a given $s_x$, changing the sign of $\tilde{s}_x$ corresponds to reversing the in-plane polarity of the  spin system. Therefore, the DM term favors a consistent alignment of in-plane and cycloidal polarities, always parallel or always antiparallel depending on the sign of $D$. This lifts the degeneracy between the two types of non-symmetry-equivalent domains --- an essential ingredient to explain why cycloids are at all observed by diffraction.

The effect of E$_{DM}$ on the wider phase diagram, away from the umklapp points, is more difficult to calculate analytically. Preliminary numerical calculations suggest that, in addition to making $\alpha$ non-zero for the umklapp phases, the DM term has the additional effect of extending the region of incommensurability --- unsurprisingly perhaps, because it promotes non-collinearity.

\section{Magnetic Domain switching under an applied electric field, Spherical Neutron Polarimetry}
\label{section:cryopad}
The simple model outlined in Sections \ref{section: NN_NNN} and \ref{section: DM} paints a picture of the \rmno physics that is rather different from that of the typical cycloidal multiferroics. The case of \Bi  demonstrates with particular clarity that the cycloidal component does not emerge independently as for \Tbun, but is induced by and has consistent polarity with the in-plane magnetic structure. The presence of an antisymmetric energy term that lifts the degeneracy between configurations with parallel and antiparallel polarities is also important, because it has a direct implication on the domain formation and on the domain switching upon application of an electric field.  In the presence of this term, if the configuration $(\alpha, \beta)$ is the stable state, the "alternate" configuration with reversed cycloids, $(-\alpha, -\beta)$ is not an extremum of the energy function. However, if $E_{DM}$ is small, in general there will be a local minimum near the $(-\alpha, -\beta)$ point, which corresponds to a metastable state. In this case, it is possible in principle to reverse one of the two polarities (in-plane or cycloidal) without affecting the other. Which of the two polarities is reversed depends on the specific coupling mechanism to the electric field; in other words, if the cycloids were entirely responsible for the electrical polarization through spin-orbit coupling, the direction of rotation of the cycloids must be reversed when the electrical polarization is reversed. Likewise, reversal of the in-plane polarity is a strict requirement of the exchange-striction model. For certain values of the parameter, the metastable minimum is very shallow or absent:  in this case, we would expect that the electric field would switch the structure between truly degenerate domains related by inversion. We can summarize the different scenarios as follows (metastable states can be "shallow" or "deep" with respect to $k_BT$):

\begin{enumerate}
\item{} \underline{FE by spin-orbit coupling - deep metastable state}.  The prediction for this scenario is that only the cycloids will switch upon reversal of the electric polarization by an applied external field. The in-plane structure should remain unaffected.
\item{}\underline{Both spin-orbit and exchange-striction contribute to FE, deep metastable state}. Here, both in-plane and cycloidal components should switch, possibly at different fields. This scenario is unlikely, as it should produce a 2-step FE hysteresis loop that is not observed experimentally.
\item{} \underline{FE by exchange-striction, deep metastable state}. Here, only the in-plane structure should switch, whereas the cycloids should not be affected.
\item{}\underline{Any mechanism, shallow metastable states}. Here, the two components always switch simultaneously, so only domains related by inversion symmetry are ever observed.
\end{enumerate}

A direct observation of the response of the magnetic structure to the reversal of an external electric field is therefore non-trivial and potentially informative about the mechanism of ferroelectricity.

Very few techniques are available to probe directly the domain structure of an antiferromagnet. Second-harmonic light generation has been used successfully to this effect in multiferroics --- see for example \cite{ISI:000178769800040}. However, this technique can only be used for "$\Gamma$-point" antiferromagnets, in which the magnetic unit cell coincides with the primitive chemical cell. In fact, the macroscopic quantity that is relevant for magnetic SHG is a third-rank, time-reversal-odd axial tensor. For non $\Gamma$-point antiferromagnets, time-reversed translations are symmetry operators. Therefore, the magnetic point group contains time reversal, and the tensor is identically zero. Another way of saying the same thing is to observe that SHG is an optical technique, and should therefore only be sensitive to zone-center ($\Gamma$-point) effects. In these more complex cases, the technique of choice to probe antiferromagnetic domains is scattering of polarized neutrons, of which Spherical Neutron Polarimetry (SNP) is a particularly powerful version. The neutron spin polarization is defined as the statistical average over the neutron beam of the expectation value of the quantum mechanical spin projection operator. An almost fully polarised ($>$ 99 \%) monochromatic neutron beam is generated by means of a Heusler crystal monochromator.  The neutron spin direction can be altered in one of three ways:
\begin{enumerate}
\item{} \underline{adiabatic rotation}:  the spins will follow a slowly-varying magnetic field;
\item{} \underline{precession}: when crossing an abrupt step change in the field direction,
the neutron spins will precess around the new magnetic field;
\item{} \underline{flip}: the neutron spins can be flipped by 180$^{\circ}$ by a special device, known as
a flipper, which also exploits neutron spin precession.
\end{enumerate}
By an appropriate combination of adiabatic rotations, precessions and flips, one can prepare the incident neutron polarization in an arbitrary direction, and also rotate the scattered polarization so that any of its components is parallel (or antiparallel) to the analysis direction. To this end, a flipper and a polarization analyzer, in this case a $^3$He spin-filter, are placed in the scattered beam in front of the neutron detector. The sample itself is held in exactly zero magnetic field. Unlike conventional neutron crystallography, spherical polarimetry does not rely on measured neutron intensities, but on intensity ratios between parallel and antiparallel settings of the final flipper/analyser pair. In practice, one can measure 9 independent "flipping ratios", by setting the incident and scattered polarization along the $X$, $Y$ and $Z$ directions (in appropriate coordinates, see \cite{Blume_PR_63}). All the elements of the resulting polarization matrix (for complete treatment see \cite{Blume_PR_63,Maleev_SPSS_63}) are sensitive to the magnetic structure factor, as is the unpolarized neutron scattering cross section. However, some of the matrix elements --- in our case the elements $P_{zx}$ and $P_{yx}$ --- carry unique information about the domain population, even if these domains are related by inversion.

We performed a SPN experiment on a \Y single crystal as a function of temperature and external electric field applied along the $b$ crystallographic direction. The results of this experiment are described in details in Ref. \cite{Radaelli_D3}. With the NPS technique, we can probe directly both domain structure and domain population for different temperatures and orientations of the external electric field.  However, on D3, there is only limited access to reflections out of the horizontal scattering plane, so the crystal mounting will dictate which region of reciprocal space is observed. For this first experiment, we chose to mount the crystal with the $b$ axis vertical, i.e., perpendicular to the scattering plane, so that only reflections with $k=0$ were accessible. This setting enabled to probe in great detail the magnetic structure within the $a-b$ plane, and to observe the effect of the domain switching on the in-plane structure. We were, however, not able to distinguish between stable and "alternate" domains. Consequently, our experiment can completely corroborate or falsify scenario (i) above.  It can also provide partial information about (ii), but it is unable to distinguish conclusively between (iii) and (iv).  The main results of our experiment are summarized in Fig. \ref{cryopad}. We applied two types of electric field-switching protocols:  cooling in an applied electric field of either polarity ($\pm 2.2 kV/cm $ ) through the N\`{e}el temperature T$_N$ down to 25 $K$ (left panel) and cycling the electric field at 35 $K$, i.e., closer to T$_N$, after polarizing the sample at 25 $K$.  For in-field cooling, we observed a complete reversal of the in-plane domain population (Fig. \ref{cryopad}, left). This clearly demonstrates that the in-plane magnetic structure is coupled to the electric field, ruling out scenario (i) above. At 35 $K$, we were able to measure a complete hysteresis loop on the off-diagonal neutron polarization matrix elements (Fig. \ref{cryopad}, right). The loop is shifted downwards with respect to the centerline, indicating that some domains are "locked" in a fixed polarization direction and cannot be reversed by the small electric field available to us. More importantly, the SNP results are in excellent agreement with measurements of the macroscopic electrical polarization performed with the same protocol, including the downward shift of the hysteresis loop and limiting values of the electrical polarization at the top and bottom of the hysteresis loop. In other words, the in-plane antiferromagnetic domain population is strictly proportional to the electrical polarization. This result indicates that, if both in-plain and cycloids contributed independently to the polarization (scenario (ii)), the latter contribution is probably very small. In a future experiment we are planning to perform an independent measurement of the cycloidal domain switching under similar conditions, by mounting the crystal with a different orientation. This will enable us to discriminate among the above scenarios in a unique way.

\section{Outlook and conclusions}
\label{section:conclusions}

\indent As we have seen in the previous sections, much progress has been made towards understanding the connections between magnetism and ferroelectricity in \rmno, thanks to the sustained effort of several groups. Neutron diffraction played a major part in this research, providing several key pieces of information. Nonetheless, much remains to be done. The most obvious gap is the lack of a definitive determination of the magnetic structures of the two ICPs. The HT-ICP is clearly the most difficult one, because the magnetic moments are small, and no solution has been proposed to date. Starting form the seventies, several groups have attempted to solve the magnetic structure of the LT-ICP, from both powder \cite{ISI:A1973P583900022,ISI:000235905700072} and single-crystal data \cite{ISI:A1981LP35700027,ISI:A1988R221000009,Kim}. Some features are common to all these refinements, and can be considered as established with good confidence. Other aspects are slowly emerging as techniques and refinements improve.  The main results to date can be summarized as follows:

\begin{enumerate}
\item{} In the LT-ICP, the modulations of the two zig-zag chains are approximately \emph{in quadrature}, rather than \emph{in phase}, as in the CP.  This emerges clearly both from the powder data \cite{ISI:000235905700072} and the more recent single-crystal data \cite{Kim}.  This is a key feature in the context of the exchange-striction model, since it explains why ferroelectricity is suppressed in the LT-ICP:  in fact, all terms in exchange-striction polar vector $E_y$ (eq. \ref{eq: ES_ICP}) contain the cosines of the phase differences between atoms on different zig-zag chains.  If the phase difference between these chains is close to $90^{\circ}$, the cosines will be small and so will be the electrical polarization.
\item{}  In the LT-ICP magnetic structure, one can recognize "patches" resembling CP domains \cite{ISI:000235905700072}.  Both polarities are equally represented, so their contribution to ferroelectricity cancel out.  These "patches" alternate with regions where the dot product between sites on the two chains is very small or zero, yielding a negligible contribution to ferroelectricity.
\item{}  The main difference between  the powder structure \cite{ISI:000235905700072} and the more recent (and accurate) single-crystal structure \cite{Kim} is the relative phase of the $b$ component of the spins.  In the latter, the $S_x$ and $S_y$ components are \emph{in quadrature} yielding a cycloidal component in the $a-b$ plane.  Here, the spins of the two chains are almost orthogonal throughout the modulated strucure. This has no impact on the exchange-striction polar vector calculation, since the two components contribute independently to $E_y$, but may provide an additional spin-orbit contribution.  We remark that if the two zig-zag chain modulation were identical except for a phase factor --- a reasonable hypothesis --- this new spin-orbit contribution would be directed entirely along the $b$ axis.
\item{} Assessing the the spin-orbit polar vector in the LT-ICP requires a level of accuracy in the magnetic structure determination that, in our opinion, has not yet been attained.  In Section \ref{section:origin}, we present the results based on the recent determination by Kim \textit{et al.} \cite{Kim}, which are not in accord with the experimental values of the electrical polarization.  We remark, however, that the structure by Kim \textit{et al.} is not yet completely satisfactory, since symmetry-equivalent atoms have widely different modulation amplitudes. We believe that these inconsistencies should be resolved before a final assessment of the relative importance of the two mechanisms can be made.
\end{enumerate}

Another aspect so far remains largely unexplored is that of rare earth magnetic ordering.  It is now known that a sizeable magnetic moment is induced on the RE by the ordering of Mn (see, for example, \cite{Beutier}).  Some aspect of the low-temperature ordering and the effect of the application of an external magnetic field have also been established \cite{ISI:000224662700076}.  However, the dramatic changes of the electrical polarization upon RE magnetic ordering have not been explained.  The study of RE with different on-site anisotropy may provide important clues to understand these effects.

In summary, we have presented an overview of the neutron diffraction results on the \rmno multiferroics, with particular emphasis on the correlations between magnetic ordering and ferroelectricity.  Solely on the basis of the refined magnetic structures, we calculate polar-vector quantities that can be directly related to the electrical polarization, for both exchange-striction and spin-orbit microscopic models of the dominant magneto-electric interactions. We also explored the underlying causes giving rise to the complex magnetic phase diagram of \rmno, and proposed a "minimal" model that can reproduce some of the observed features, whilst highlighting the importance of antisymmetric exchange in stabilizing phases of consistent polarity.  Finally, we illustrated what in our opinion are the missing "pieces of the puzzle" required to understand fully the phenomenology of these remarkable materials.

\bibliographystyle{iopart-num}
\providecommand{\newblock}{}

\pagebreak

\pagebreak

\pagebreak

\pagebreak----

\begin{figure}
\includegraphics[scale=1.6]{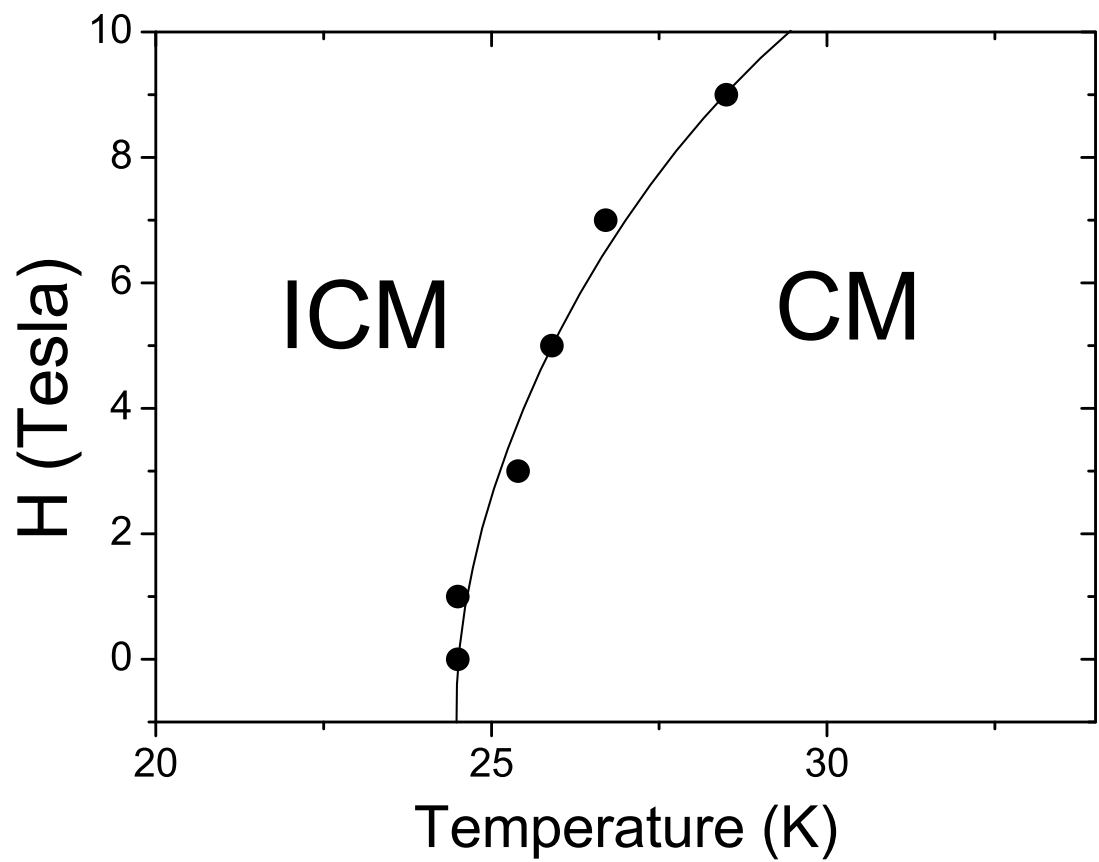}
\caption{H-T phase diagram of the commensurate (CM) to
incommensurate (CM) phase transition in \Tb\. The points have been
calculated by a fitting procedure (see text for details). The solid
line is a guide to the eyes.}
\label{figurefield2}
\end{figure}

\pagebreak

\begin{figure}
\includegraphics[scale=1.6]{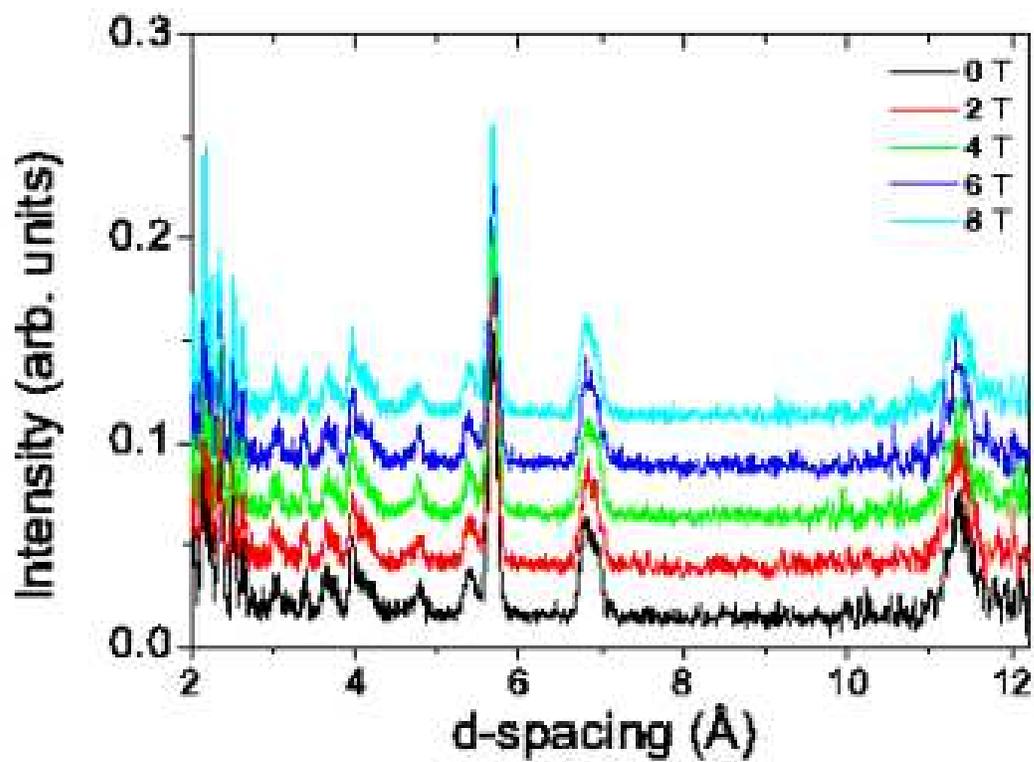}
\caption{Time-of-flight diffraction patterns of YMn$_2$O$_5$ at 1.6K under magnetic fields between 0 and 8 Tesla.}
\label{figurefield3}
\end{figure}

\pagebreak

\begin{figure}[]
\includegraphics[scale=1, angle=-90]{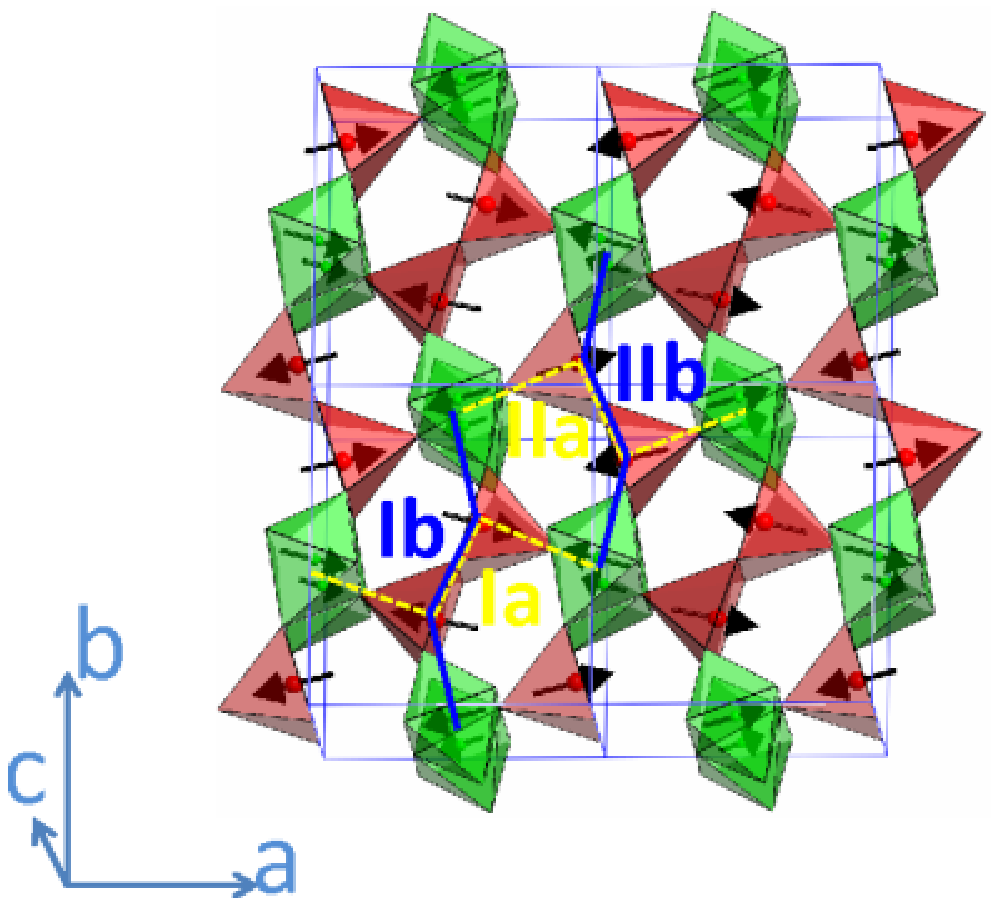}
\\
\\
\includegraphics[scale=1, angle=-90]{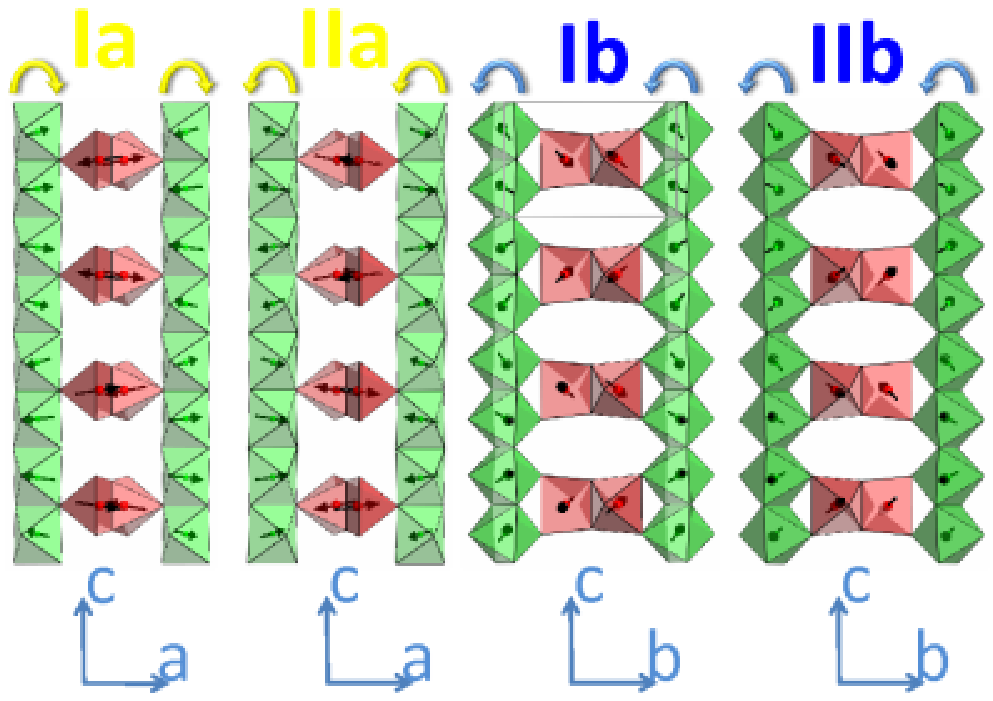}
\caption{Representation of the commensurate magnetic structure at 25K. The model is derived from analysis of single-crystal neutron diffraction data. Magnetic ions are represented with the same color scheme defined in reference \cite{vecchini} . Magnetic moments are represented by black arrows. Top panel: Magnetic configuration in the ab-plane (2x2 unit cells) showing the presence of nearly antiferromagnetic zig-zag chains (see text for details). Several $Mn^{4+}$-$Mn^{3+}$-$Mn^{3+}$-$Mn^{4+}$ fragments in the structure are labelled by roman numbers (Ia and IIa along the a-axis and Ib,IIb along the b-axis). Bottom panel: Magnetic structute projected in the ac- and bc- planes. Configurations within each fragment (Ia, IIa, Ib,IIb) is shown for 4 unit-cells along the c-direction. The curled arrows at the top of each magnetic chain represent the rotation direction of the cyloidal modulation. For clarity, the size of the magnetic moment has been doubled for representing modulations in the bc-plane (Ib,IIb) }
\label{magnetic}
\end{figure}

\pagebreak

\begin{figure}
\includegraphics[scale=1, angle=-90]{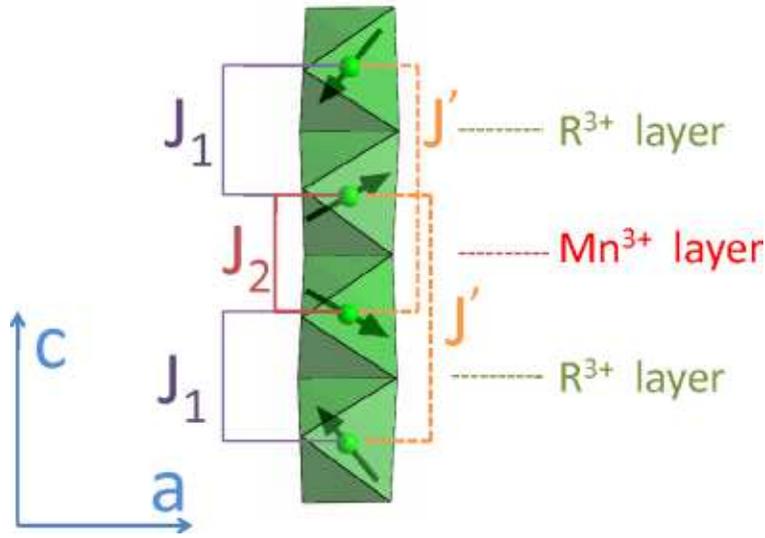}
\caption{ Labeling scheme and relevant interactions for the linear chain model discussed in this section.  One of the possible spin ordering modes is shown.}
\label{exchangealongchains}
\end{figure}

\pagebreak

\begin{figure}
\includegraphics[scale=0.60, angle=-90]{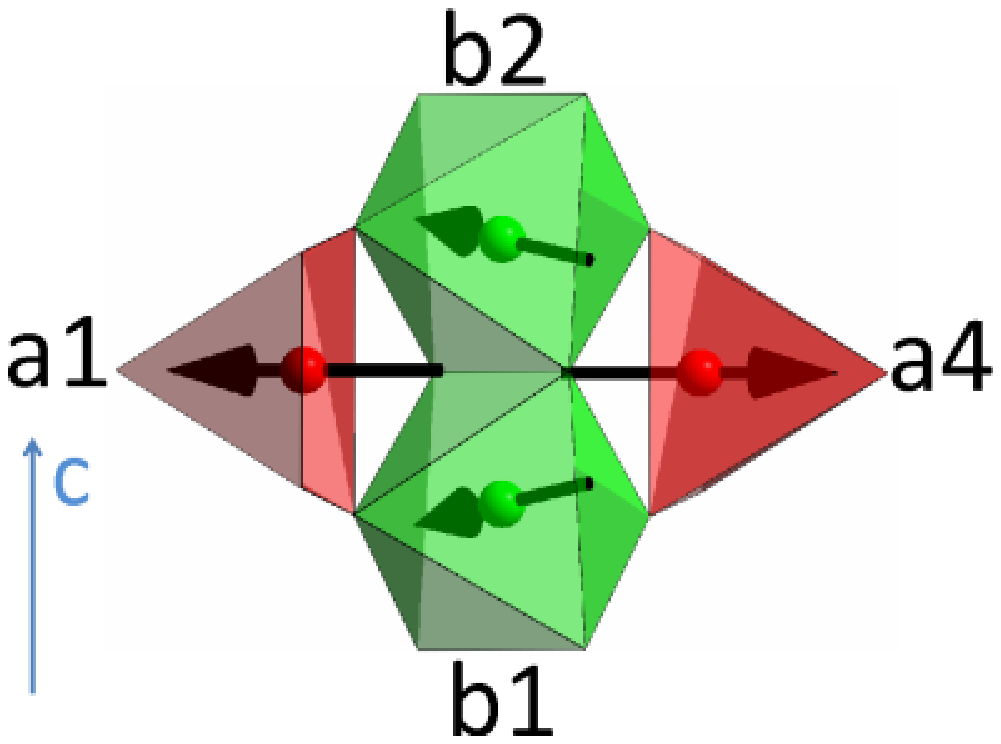}
\includegraphics[scale=0.60, angle=-90]{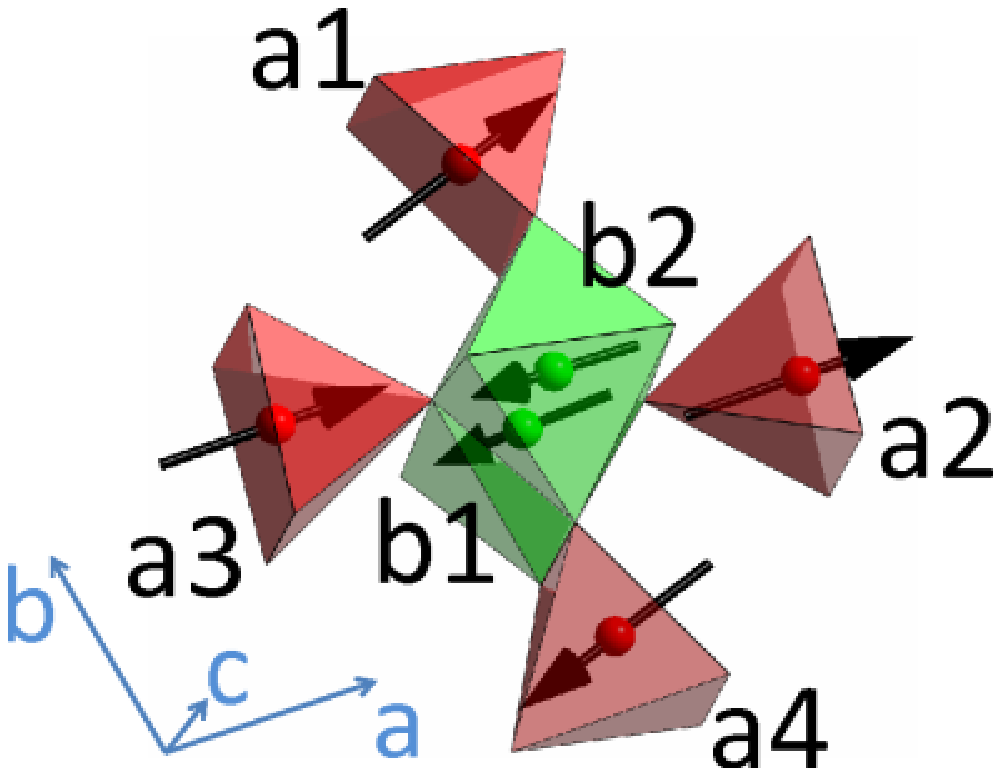}
\caption{Magnetic site labeling scheme employed for the calculation of the DM energy, from \cite{vecchini})}
\label{figureDM}
\end{figure}

\pagebreak

\begin{figure}
\includegraphics[scale=0.8, angle=-90]{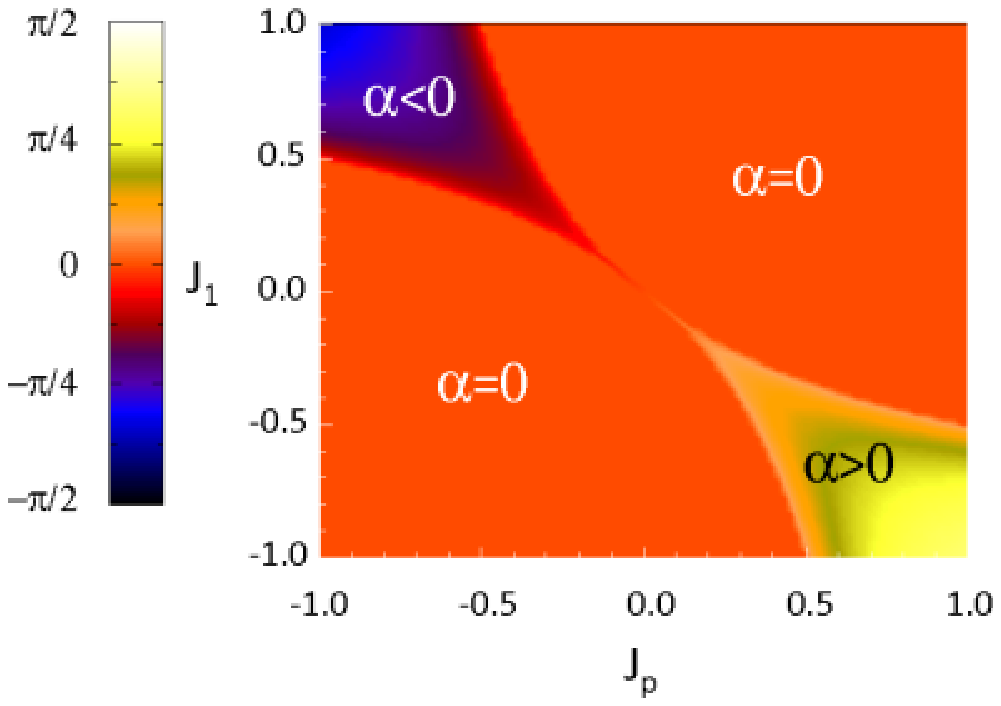}
\includegraphics[scale=0.8, angle=-90]{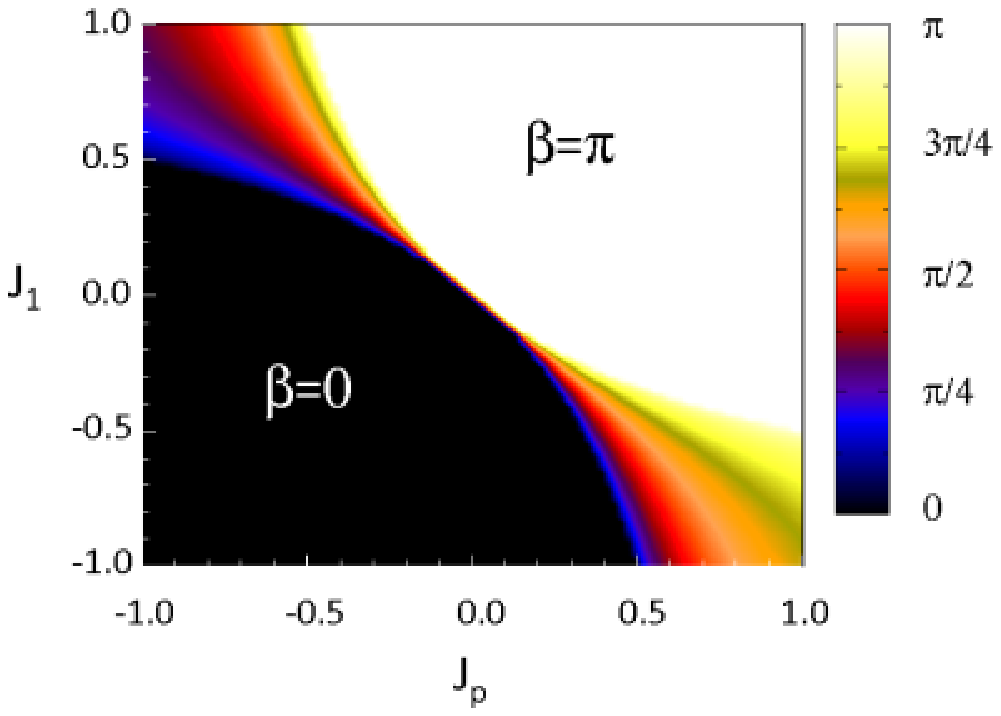}
\caption{Phase diagram for the $J_1-J_2-J^{\prime}$ model for $J_2=-1$ and $\Gamma=0$ (isotropic case).  Areas in solid colors indicate commensurate phases, whereas a color gradation indicates incommensurability.  The angle $\alpha$ is twice the phase offset between sites $S_1$ and $S_2$, whereas $\beta$ is the propagation vector $q_z$ times $2\pi$ (see text).}
\label{Fig:phase_diagram}
\end{figure}

\pagebreak

\begin{figure}
\includegraphics[scale=0.8]{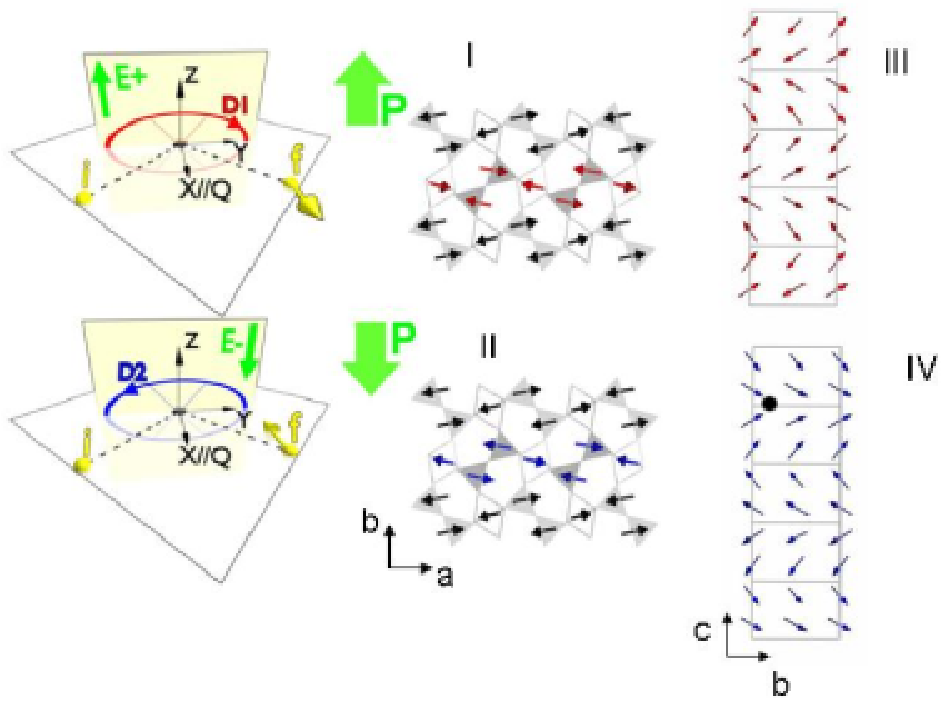}
\includegraphics[scale=1.5]{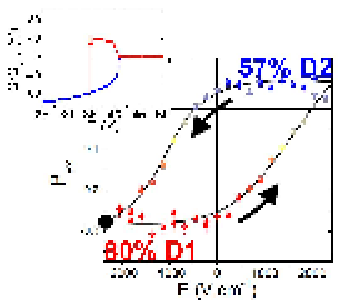}
\caption{\textbf{Left}:  Schematic representation of the neutron spherical polarimetry experiment for the two domains, here shown in the idealized case of an unpolarized incident beam.  The real (or imaginary) component of the magnetic structure factor projection $\textbf{\textrm{M}}^{\bot}_{hkl}$ rotate clockwise (counter clockwise) for Domain \textbf{I} (Domain \textbf{II}), creating a spin polarization of opposite signs for the scattered neutron.  The direction of the applied electric field is also indicated.  \textbf{Right}: Magnetic structures of YMn$_2$O$_5$ for the different domain configurations, projected on the $ab$ plane (\textbf{I} and \textbf{II}) and the $bc$ plane (\textbf{III} and \textbf{IV}). Small arrows represent magnetic moments. The observed domain switching mechanism is represented by the inversion (change from the red to the blue) in the central chain (between configurations \textbf{I} and \textbf{II} in the $ab$  plane). \label{Fig: polarimetry}Right panel: Partial hysteresis loop measured on the created neutron polarization element $P_{yx}$  for the $-\frac{1}{2}\, 0\, -\frac{7}{4}$  Bragg peak of an YMn$_2$O$_5$ crystal, warmed to 35 K after previous cooling to 25 under a negative -2.2 kV/cm electric field.  \textbf{inset}:  integrated pyroelectric currents measured on a 0.5 mm thick YMn$_2$O$_5$ crystal of the same batch on cooling to 25 K in a negative -2.0 kV/cm electric field (bottom/blue curve), followed by warming to 35 K and switching to a positive +2.0 kV/cm electric field (top/red curve).  The data are normalized to the fully saturated value at 25 K.  Both hysteresis bias and the switching ratios are in very good agreement with the neutron data.}
\label{cryopad}
\end{figure}

\end{document}